\documentclass[acmsmall,screen,nonacm]{acmart}
\usepackage{tabularx}
\usepackage{multirow}
\usepackage{flafter}
\usepackage{placeins}
\usepackage{needspace}
\usepackage{algorithm}
\usepackage{algorithmic}
\usepackage{xspace}
\usepackage{mathpartir}

\makeatletter
\if@ACM@nonacm
  \let\@ACM@checkaffil\relax
\fi
\makeatother

\newcommand{\method}{\textsc{Synapse}\xspace}
\newcommand{\skir}{\textsc{SkIR}\xspace}
\newcommand{\skast}{\textsc{SkAST}\xspace}
\newcommand{\Execute}{\textsc{Execute}\xspace}
\newcommand{\Diagnose}{\textsc{Diagnose}\xspace}
\newcommand{\Propose}{\textsc{Propose}\xspace}

\newcommand{\chong}[1]{#1}
\newcommand{\juan}[1]{#1}
\newcommand{\cue}[1]{``#1''}

\newcommand{\parhead}[1]{\par\textit{#1.}}

\makeatletter
\newtheoremstyle{acmplaintight}%
  {.25\baselineskip\@plus.1\baselineskip\@minus.1\baselineskip}%
  {.25\baselineskip\@plus.1\baselineskip\@minus.1\baselineskip}%
  {\@acmplainbodyfont}{\@acmplainindent}{\@acmplainheadfont}{.}{.5em}%
  {\thmname{#1}\thmnumber{ #2}\thmnote{ {\@acmplainnotefont(#3)}}}
\newtheoremstyle{acmdefinitiontight}%
  {.25\baselineskip\@plus.1\baselineskip\@minus.1\baselineskip}%
  {.25\baselineskip\@plus.1\baselineskip\@minus.1\baselineskip}%
  {\@acmdefinitionbodyfont}{\@acmdefinitionindent}%
  {\@acmdefinitionheadfont}{.}{.5em}%
  {\thmname{#1}\thmnumber{ #2}\thmnote{ {\@acmdefinitionnotefont(#3)}}}
\renewenvironment{proof}[1][\proofname]{\par
  \pushQED{\qed}%
  \normalfont \topsep.25\baselineskip\@plus.1\baselineskip\relax
  \trivlist
  \item[\@proofindent\hskip\labelsep
        {\@proofnamefont #1\@addpunct{.}}]\ignorespaces
}{%
  \popQED\endtrivlist\@endpefalse
}
\makeatother
\theoremstyle{acmplaintight}
\newtheorem{theorem}{Theorem}[section]
\newtheorem{lemma}[theorem]{Lemma}
\newtheorem{proposition}[theorem]{Proposition}

\theoremstyle{acmdefinitiontight}
\newtheorem{definition}[theorem]{Definition}

\theoremstyle{acmplaintight}

\newcommand{\papertablestyle}{%
  \normalfont\small%
  \setlength{\tabcolsep}{8pt}%
  \renewcommand{\arraystretch}{1.15}%
}

\AtBeginDocument{%
  }

\begin{document}

\title[\method: Optimizing Agent Skills through Program Repair]{\method: Optimizing Natural-Language Agent Skills through Program Repair}

\author{Lijuan Li}
\affiliation{\institution{Nanjing University}}

\author{Zhenpeng Chen}
\affiliation{\institution{Tsinghua University}}

\author{Chenxi Zhang}
\affiliation{\institution{Xidian University}}

\author{Yihao Huang}
\affiliation{\institution{East China Normal University}}

\author{Yang Liu}
\affiliation{\institution{Nanyang Technological University}}

\author{Chong Wang}
\affiliation{\institution{Nanyang Technological University}}

\renewcommand{\shortauthors}{Li et al.}

\begin{abstract}
Large language model (LLM) agents can improve task performance by optimizing reusable natural-language skills without updating model parameters.
However, \chong{the execution feedback that guides this optimization} identifies failed tasks without directly revealing which instructions should be modified, which other instructions a fault involves, or which other tasks an edit may break.
\chong{For programs, fault localization, dependence analysis, and regression test selection answer these three questions.}
Our key insight is that, although a skill is not a program, the procedure it describes can be partly recovered as one, and that this \chong{partial} program can serve as an editable view of the skill.
We propose \textbf{\method}, which casts skill optimization as generate-and-validate repair over \chong{this view}.
\chong{A deterministic compiler that, like an island grammar, recognizes only the constructs of interest} translates a skill into \skir, a source-linked intermediate representation that exposes operations, conditions, and data references and keeps every other requirement in the text.
\method localizes faults from traces of repair tests\chong{, the tasks with reference answers that guide the search,} and relates each fault to neighboring instructions through control-flow and dependence relations derived from \skir.
\chong{It admits only typed \skir mutations, which it writes back to} \chong{the mapped spans of the instructions they target and recompiles.
It screens each patch on related repair tests, including previously passing ones, before validation decides whether to accept it.}
Whatever the \chong{LLMs that execute the skill and propose patches} return, \chong{every accepted patch yields a well-formed skill and leaves the source span of every untargeted instruction unchanged, and if the execution agent is one-hop local, the selected tests include every test that an edit can affect}.
\chong{On two single-turn question-answering benchmarks, DocVQA and LiveMath, with three execution models, \method improves task success rates over SkillOpt by \juan{3.74--4.54} percentage points on DocVQA and 2.40--4.00 on LiveMath and outperforms Trace2Skill and SkillBoost in all six settings of benchmark and execution model.}
\chong{In an ablation, replacing} \skir-based repair and \chong{its} test screening with constrained text editing screened only on failing tests reduces DocVQA success rates by \juan{2.40} and 2.67 percentage points for GPT-5.6 Luna and GPT-6 Sol, respectively.
In two recorded DocVQA repair runs, control-flow and dependence relations add related repair tests to the screening of 23 of 37 patches\chong{, and in one repair, a test added only through a control-flow relation exposes a regression that coverage of the edited instruction misses}.
\end{abstract}



\maketitle

\section{Introduction}\label{sec:introduction}

LLM agents increasingly perform complex tasks by planning actions, invoking tools, and interacting with external environments~\cite{react2023,schick2023toolformer,wang2024voyager,yang2024sweagent,wang2024agentsurvey}, yet many tasks require procedural knowledge that is difficult to encode in the model itself.
Agent Skills, introduced by Anthropic~\cite{anthropic2025skills} and released as an open standard~\cite{agentskills2025spec}, supply such knowledge externally.
A skill document tells an agent which steps to follow, when \chong{they apply}, what information to check, and how to return an answer.
Skills have been adopted by many agent products~\cite{agentskills2025spec}\chong{, and as of October 2026,} the cross-agent directory skills.sh ranks about 1.5 million skills by installation~\cite{skillssh2026}.

Because skills are written in natural language, often by community authors or by LLMs, their quality is hard to guarantee.
On SkillsBench, model-generated skills perform worse than no skill, whereas curated skills help~\cite{skillsbench2026}; skills that appear relevant can still cause a task to fail~\cite{dong2026harmful}; and in a study of three skill libraries, regressions offset 59\% of the gains that skills bring~\cite{tank2026regression}.
Skills therefore need to be improved from their effects on task execution.
\chong{Recent work on \emph{skill optimization} revises a skill from tasks with reference answers so that an agent following it answers unseen tasks of the same kind correctly more often~\cite{trace2skill2026,skillopt2026,lin2026skillboost}.}

\chong{Improving a skill, however, takes more than rewriting its text. Consider a DocVQA skill whose \emph{value-span instruction} tells the agent to return only the requested value span and to omit units unless the question asks for them. Asked for the total turnover in a document that prints \cue{41,809.82~crores}, an agent that follows the skill answers \cue{41,809.82} and drops the scale word \cue{crores}, which multiplies the amount by ten million. A repair must answer three questions. Which instruction should change, and what kind of change does it need? Which other instructions does the failure involve, such as the condition under which this instruction applies? Which passing answers might the edit break, such as a receipt amount in words? Existing optimizers answer them only in part.}

\chong{These optimizers differ in how they find and make an edit.}
\emph{Trace-driven} methods revise or patch skill documents from execution trajectories~\cite{trace2skill2026,skillopt2026,skillgrad2026}, \emph{attribution-guided} methods first identify the sections, skills, or fields responsible for failures~\cite{reskill2026,lin2026wml}, \emph{structure-aware} methods apply typed operators to steps, rules, or workflow paths~\cite{ma2026sepo,sun2026graphskillevo}, and \emph{library-level} methods expand and consolidate skill collections~\cite{skillcommit2026}.
\chong{Measured against the three questions, they share three limitations.}
\textbf{(L1) Imprecise edit targets.}
\chong{Trace-driven and attribution-guided methods express the edit itself as rewritten text, and structure-aware methods, including the concurrent GraphSkillAA~\cite{skillaa2026}, apply typed edits to a separate structured representation of the prompt or skill~\cite{ma2026sepo,sun2026graphskillevo}, which does not determine which span of an existing skill an edit replaces.
\chong{To our knowledge, no skill optimizer ties an edit to a typed element of the existing skill, namely the instruction it modifies, the kind of change it makes, and the span it replaces. Without this tie, an optimizer has no declared target and kind of change against which to check an edit before running the edited skill, such as a rewrite confined to the value-span instruction in the turnover example.}} \textbf{(L2) Implicit instruction dependences.}
\chong{Structure-aware and library-level methods} organize instructions into steps, rules, workflow paths, \chong{or typed skill-graph edges}~\cite{ma2026sepo,sun2026graphskillevo,skillaa2026} or manage relationships among skills~\cite{skillcommit2026}\chong{, and SkVM extracts the dependencies between steps to run independent steps in parallel~\cite{chen2026skvm}}.
\chong{To our knowledge, no skill optimizer derives the control and data dependences among the instructions of an existing skill from its text, so an optimizer must infer which other instructions a failure involves.
In the turnover example, the condition under which the value-span instruction applies is stated in a separate clause, so the repair model must also see a clause outside the instruction.}
\textbf{(L3) \chong{Untargeted} regression checks.}
\chong{Existing methods check an edit against the motivating failure~\cite{reskill2026}, source tasks~\cite{skillcommit2026}, a fixed task set~\cite{skillopt2026,lin2026skillboost}, or the examples that an update affects in a skill graph~\cite{skillaa2026}. An edit can also change the answer to a task whose trace never covers the edited instruction, through a related instruction that the trace does cover.
In the turnover example, the receipt is such a task, since its trace covers a neighboring condition but not the value-span instruction.
To our knowledge, no skill optimizer selects the tasks to recheck through dependences recovered from an existing skill's text, so such a regression is caught only if that task happens to be checked.}

\chong{For programs, generate-and-validate repair has an established answer to each of these limitations.}
Fault localization~\cite{jones2002spectrum} and repair operators~\cite{legoues2012genprog} determine where and how a patch changes \chong{a} program (L1), dependence analysis reveals which statements relate to a fault~\cite{aho2006compilers,ferrante1987pdg} (L2), and regression test selection identifies the tests that a change may affect~\cite{rothermel1997safe} (L3).
These techniques operate on programs, however, and a skill is a document.
Structured skill representations~\cite{liang2026ssl,xu2026skillsmith} and work that treats prompts as programs~\cite{liang2025promptsprograms,sharma2025promptpex,schnabel2024sammo} show that skills carry program-like structure, \chong{but none recovers this structure from an existing skill's text and repairs the skill through it.}

Our key insight is that the procedure a skill describes can be partly recovered as a program that serves as an editable view of the skill.
Like an island grammar~\cite{moonen2001island}, which parses only the constructs of interest in a text and skips the rest, a deterministic compiler can extract a skill's operations, conditions, and data references and link each \chong{of them} to its span in the document.
\chong{Every other requirement stays in the text of its instruction, where the analyses ignore it but the agent that follows the skill still reads it.}
We call the result a \emph{partial program view}.

\chong{Even on this view, two properties of skills keep the program-repair techniques from transferring directly.}
First, instructions resist reliable parsing, so the recovered structure is incomplete, and an edit made on it must still land in the authored text.
\chong{In the turnover example, the exception about units is not a recognized element, yet it is the text that the repair must change.}
\chong{Second, the interpreter of a skill, its \emph{execution agent}, is an LLM agent.}
\chong{The interpreter cannot be instrumented, so which instructions it used is known only from its report}, its answers vary across requests, and no static relation predicts the effect of an edit, so edits must be checked by running the skill.

\chong{We present \textbf{\method}, which casts skill optimization as generate-and-validate repair over the partial program view of a skill.
\method compiles a skill into \skir, a source-linked intermediate representation, and derives its control-flow and dependence graphs.
It localizes each failing repair test, a task with a reference answer that guides the search, to the instructions that the execution agent reports using.
A second LLM, the repair model, sees each suspicious instruction together with its directly related instructions and may propose only typed mutations of \skir elements, such as an instruction, a condition, or a dependency (L1, L2).
\method writes each patch back to the mapped spans of the instructions it targets and recompiles the result, so every patch is checked for well-formedness before any test runs.
To catch regressions that coverage alone misses (L3), it screens a patch on the repair tests that the reported coverage and the control-flow and dependence relations select.
It accepts a patch only if it raises the pass count on a separate validation suite.}

\chong{Because a deterministic host checks every LLM result, the properties it enforces in each iteration hold whatever the LLMs return.}
\chong{For programming languages, the approach repairs a language with a partial grammar and an LLM interpreter, through a recovered view whose edits stay within their spans and whose test selection is safe under a weaker locality assumption than coverage needs.}

\chong{Following SkillOpt~\cite{skillopt2026}, we evaluate \method on DocVQA~\cite{mathew2021docvqa}, which asks questions about documents, and LiveMath~\cite{he2026livemathematicianbench}, which poses multiple-choice mathematical problems, with one repair model, three execution models, and held-out test tasks.
The optimized skills improve \juan{test success rates} over the initial skill by 4.28--14.71 percentage points on DocVQA and 4.80--9.60 on LiveMath, and exceed SkillOpt~\cite{skillopt2026}, Trace2Skill~\cite{trace2skill2026}, and SkillBoost~\cite{lin2026skillboost} in all six \chong{settings of benchmark and execution model}, \chong{by 1.60--4.54 points over the strongest in each setting}.
\chong{In an ablation with two execution models, replacing \skir-based repair and its screening with constrained text editing lowers success rates by 1.60--3.20 points.}
\chong{In two recorded DocVQA repair runs, control-flow and dependence relations add related repair tests to the screening of 23 of 37 patches\chong{, and in the turnover repair, a test selected only through a control-flow relation exposes a regression that coverage of the edited instruction misses}.
Only the final skill of \method answers the turnover question with \cue{41,809.82~crores}.}}

\begingroup\csname @beginparpenalty\endcsname=10000
In summary, this paper makes the following contributions:
\begin{itemize}
\item \textbf{Partial program views of skills.} We introduce \skir,
a source-linked intermediate representation that a deterministic \chong{island-grammar-style} compiler recovers from skills, with \chong{write-back and recompilation checks} that make it editable. \chong{Whatever the LLMs return, accepted patches} leave every untargeted instruction's source span unchanged.
\item \textbf{Program repair for an LLM-interpreted language.} We
adapt fault localization, dependence analysis, and regression test
selection to skills, where \skir relations choose \chong{a repair's context and screening tests. Selection through these relations is safe if the execution agent is one-hop local, a weaker assumption than coverage-only selection needs.}
\item \textbf{Evaluation.} On DocVQA and LiveMath with three
execution models, \method achieves higher success rates than SkillOpt,
Trace2Skill, and SkillBoost in all six settings. We ablate \chong{\skir-based repair with its test screening, analyze how localization, write-back, and test selection operate,} and will release the implementation, skills, and repair logs.
\end{itemize}
\endgroup

\section{\method in a Nutshell} \label{sec:background}

\begin{figure}[!t] \centering \includegraphics[width=\linewidth]{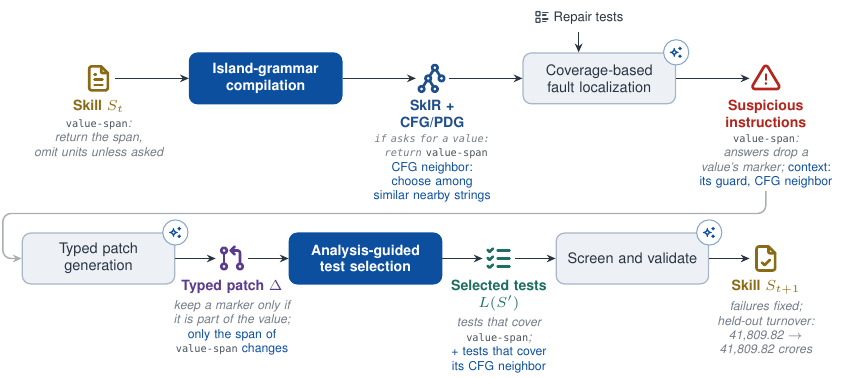} \caption{One repair iteration of \method.
Blocks are steps and icons are the artifacts they produce, and the grey italic text under each artifact shows it for \chong{the turnover example}.
The two solid blocks are the program-analysis steps, \chong{which call no LLM.
One compiles the skill into a partial program view, and the other selects the tests that screen a patch through the \chong{control-flow and dependence relations} of that view.}
\chong{\skir, linked to the skill text by its source map, forms that view, shown with its control-flow graph (CFG) and program dependence graph (PDG).}
Steps marked \protect\raisebox{-.3ex}{\protect\includegraphics[height=1.6ex]{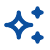}} call an LLM.
The execution agent runs the skill to localize failures and to screen and validate patches, and the repair model \chong{forms repair hypotheses and} generates typed patches.
\chong{Each patch is written back to the skill text and recompiled before its tests are selected. $S_t$ is the current skill, $\Delta$ a typed patch, $L(S')$ the tests selected for the patched skill $S'$, and $S_{t+1}$ the next skill.}
A patch is accepted only if \chong{the patched skill} passes more validation tests \chong{than the current skill}.}
\label{fig:overview} \Description{A two-row method flow of blocks (steps) and icons (artifacts).
Row one: the skill is compiled by island-grammar compilation into SkIR with its CFG and PDG; coverage-based fault localization, fed by the repair tests and using an LLM, yields suspicious instructions, in the example the value-span instruction, whose failing answers drop a marker that belongs to the value.
Row two: typed patch generation, using an LLM, produces a typed patch that keeps such a marker; analysis-guided test selection adds the tests that cover the instruction's CFG neighbor; screening and validation, using an LLM, yield the next skill, which answers the held-out turnover question with 41,809.82 crores instead of 41,809.82.}
\end{figure}

\method recovers a partial program from a skill and repairs the skill the way a generate-and-validate tool repairs code (Figure~\ref{fig:overview}).
\chong{We first show what program structure a skill leaves implicit and then follow one repair of the turnover example through the steps of the figure.}

\subsection{Implicit Program Structure in Skills} \label{sec:skill-background}

\chong{A skill resembles a program but leaves its structure implicit.
Its instructions describe operations, the conditions under which they apply, the information they use, and the form of the answer, often several in one instruction, as structured skill representations make explicit~\cite{liang2026ssl,xu2026skillsmith}.
For example, one instruction may say to return a selected value and when to omit units.
No syntax marks an operation, a condition, or a value.
A condition stated in one instruction may govern an operation described in another, and a value may be referred to by different phrases.}
Textual order gives the sequence of steps, but neither it nor \chong{authors' labels reveal} which condition governs which operation or which instruction supplies a value that another uses.

\chong{Although no dedicated syntax marks these elements, document conventions mark part of this structure.}
\chong{A skill is a Markdown document, in which headings delimit} sections, list items and paragraphs delimit instructions, and inline code marks the files, fields, and values that instructions refer to.
Procedural prose adds conventional forms.
An applicability clause such as \cue{When to use} states when the next instruction applies, a conditional clause such as \cue{if~\ldots} or \cue{provided that~\ldots} restricts the action of its instruction, and an imperative verb names that action.
\chong{Structure stated in these forms can be read off the text, whereas structure stated otherwise remains implicit.}

\subsection{Illustrative Example} \label{sec:why-representation}

\chong{To see what a repair of the turnover failure needs, consider the initial DocVQA skill of our experiments, a Markdown document of instructions for answering questions about scanned documents.
A separate \cue{When to use} clause placed before the value-span instruction states when this instruction applies, not when the skill as a whole is invoked, and restricts it to questions that ask for a value, count, page number, date, or graph reading.
The action of the instruction reads}
\begin{list}{}{\setlength{\leftmargin}{1.5em}\setlength{\rightmargin}{0.5em}}\item[]
\small
return only the requested value span; omit nearby labels, category
names, units, or explanatory words unless the question explicitly asks for
them.
\end{list}
\chong{The turnover question of the introduction is held out, and the initial skill answers it with \cue{41,809.82}.
The numeric text is recognized correctly, but the value-span instruction excludes a word needed to express the requested amount.
Because \cue{crores} is printed as part of the selected value and sets the scale in which the amount is expressed, keeping the value's unit here means keeping this word.
The final skills selected by SkillOpt~\cite{skillopt2026}, SkillBoost~\cite{lin2026skillboost}, and Trace2Skill~\cite{trace2skill2026} also answer \cue{41,809.82}.}

\chong{A repair must therefore distinguish a marker printed with the selected value, such as a unit or scale word, from one supplied by a header, a separate column, or the question.
To draw this distinction without breaking other answers, the repair model needs the instruction to edit, the condition under which it applies, and passing tests that delimit the change.
The partial program view connects the instruction to the other two.
The instruction's return operation is governed by its \cue{When to use} condition and is adjacent in the control-flow graph (CFG) to a condition for choosing among similar nearby strings.
Because an edit to one step can change how the agent handles the steps next to it, tests that cover this neighboring condition can check an edit to the return operation even when the agent does not report using that operation on them.}

\subsection{How \method Repairs the Skill} \label{sec:overview-walkthrough}

\chong{Figure~\ref{fig:overview} shows the repair iteration that this subsection follows for the value-span instruction.}

\parhead{Compile and analyze} \chong{The repair needs an editable location, its context, and a neighborhood from which to select tests, so \method compiles the skill into \skir and derives its graphs.}
The value-span instruction yields a return operation, linked by the source map to its span in the Markdown document, while the exception about units remains part of the instruction's text.
Its \cue{When to use} clause yields a branch condition that governs this operation, and the \chong{control dependence graph (CDG)} records the governing relation.
The CFG also links the return to the condition \chong{\cue{When several nearby strings look similar}}.

\parhead{Localize} \chong{To decide which instruction to change, \method starts not from the turnover question, which is held out, but from failing repair tests whose answers likewise omit markers printed with the requested values.
On each test, the execution agent reports which instructions it used.
These reports, the recorded coverage of the tests, point to the value-span instruction, which becomes a suspicious location with a source span (L1).
From the blanket requirement to omit units, the repair model hypothesizes that the returned span excludes markers of the requested value itself, not only markers supplied by the context.}

\parhead{\chong{Patch and write back}}
\chong{To change the instruction without breaking the answers around it, the request for a patch supplies the text of the instruction together with these two conditions and passing tests that constrain the intended change (L2).
From this request and the hypothesis, the repair model proposes} a typed patch that
rewrites the text of the value-span instruction (L1). \chong{The host offers this operator only at an instruction with a mapped source span, so before any test runs it knows that the patch replaces exactly this span.} It replaces the unit-omission requirement by distinguishing markers integral to the value from markers supplied only by context.
\begin{list}{}{\setlength{\leftmargin}{1.5em}\setlength{\rightmargin}{0.5em}}\item[]
\small
Return only the requested value span. Preserve a currency symbol or
measurement unit when it is visibly part of the selected value text or the
same filled value field and semantically completes that value; otherwise
omit nearby labels, category names, explanatory words, and symbols or units
shown only in the question, a header, a separate column, unrelated text, or
general context.
\end{list}
\chong{Writing the patch back replaces the span of this instruction and leaves every other instruction's source span unchanged, and recompilation checks that the new \skir is well formed.}

\parhead{Select tests and validate} \chong{Addressing L3, \method screens the patch on more tests than the repair model names with it.
The named tests are the failures that the patch should fix, control tests that pass, and boundary tests at the edge of the intended change.
\method adds the tests whose recorded coverage includes the edited instruction and, through the CFG and dependence relations, the tests that cover related instructions.
One of them, a receipt test about a written-out monetary amount, enters screening only through the CFG relation.
Its recorded coverage maps to the neighboring condition for choosing among similar strings but not to the edited instruction, and the repair model did not name it.
Under the patched skill, its previously correct answer gains a comma after \cue{Pounds} and is scored incorrect.
Because the patch still repairs the two failing repair tests that motivated it and keeps its control tests passing, screening keeps it, and because of the regression, it ranks behind any patch that fixes its failures without one.
Validation on the separate suite then decides, and the patched skill passes more validation tests than the current skill, is accepted, and remains the final selected skill.
Because \cue{crores} is printed as part of the selected value text, the revised instruction keeps it as a marker of the value itself, and the final skill returns \cue{41,809.82~crores} on the held-out turnover question.}

\section{Skill Optimization with \method} \label{sec:method}

\chong{The repair techniques of the introduction assume a complete program and a deterministic interpreter, whereas \method works on an incomplete view of a skill whose interpreter is an LLM.
This section formalizes skill optimization as generate-and-validate repair, motivates each step of an iteration under these conditions, and closes with the properties that hold whatever the LLMs return.}

\subsection{Skill Optimization as Generate-and-Validate Repair} \label{sec:problem}

\chong{We first formalize how an agent executes a skill and how its answers are judged, and then state skill optimization as repair over these executions.}
A skill is executed by an \emph{execution agent} $E$, \chong{which we treat as the interpreter of the skill $S$.
The agent is built on an LLM with fixed parameters, the \emph{execution model}, and reads $S$ to answer a task.}
Changing $S$ is the only way to change how $E$ answers.
\chong{The answers of $E$ are judged by tests.}
A test $i$ pairs a task input $x_i$ with a reference answer, and a fixed scoring rule, the oracle, decides whether the answer that $E$ returns for $x_i$ under $S$ passes.
For a suite $D$ of tests, $J_D(S)$ is the fraction of $D$ that $S$ passes.
\chong{Because the same input can receive different answers~\cite{ouyang2025nondeterminism}, $J_D(S)$, like every outcome in this section, refers to a recorded run.}
An incorrect answer may expose a fault in $S$, but it may also come from a reading or reasoning error of $E$ that no change to $S$ removes.
\chong{Skill repair seeks changes to $S$ that remove \chong{faults in $S$}.}

\begin{definition}[Skill repair] Given an initial skill $S_0$ and disjoint suites of repair tests $D_{\mathrm{rep}}$ and validation tests $D_{\mathrm{val}}$, skill repair seeks a skill $S$, obtained from $S_0$ by a sequence of patches, that answers unseen tasks of the same kind correctly more often than $S_0$.
Since \chong{this rate} cannot be observed, the repair tests guide the search and the validation tests decide whether each patch is accepted.
A third suite $D_{\mathrm{test}}$\chong{, held out, only estimates the final skill's pass rate on unseen tasks}.
\end{definition}

\chong{The repair tests play the part of the test suite in generate-and-validate repair~\cite{legoues2012genprog,qi2015test}, and those that a skill $S$ fails, written $F(S)$, point to the faults that a patch should remove.
As in that setting, the repair tests are} an incomplete specification, and a patch that makes them pass may still fail on other tasks~\cite{smith2015overfitting}.
\method therefore checks each patch on tests beyond the failure that motivated it and then on a separate validation suite.

\parhead{One repair iteration} \chong{\method repairs a skill iteratively, and each iteration follows the generate-and-validate loop.
Algorithm~\ref{alg:repair} turns one skill version $S_t$ into the next.
It compiles $S_t$ with the compiler $\mathcal{C}$ into its \skir $I_t$ and source map $M_t$ and records a trace for each repair test (lines~1--2), asks a repair model $O$, also an LLM, for typed patches that it writes back and recompiles (lines~3--8), and screens and validates the patched skills, of which it accepts at most one (lines~9--13).
An accepted patch yields $S_{t+1}$, and after a fixed number of iterations the final skill is evaluated on $D_{\mathrm{test}}$.}
The execution agent and the repair model are reached only through three procedures with different roles.
\Execute runs the skill, \chong{as running a test does in program repair, and is how \method observes} the behavior of the skill, both to localize faults and to validate patches.
\Diagnose and \Propose call the repair model, which plays the role of the patch generator.

\begin{definition}[Execution and repair procedures] \emph{Execution.} $\Execute(S,x)$ runs the execution agent $E$ on task input $x$ with the skill $S$, extended by a deterministic address contract that labels its instructions and \chong{the data dependencies among them}, and returns the answer of $E$ together with the addresses that $E$ reports using.
\emph{Repair.}
The repair model $O$ is called through two procedures. $\Diagnose(\Lambda)$ receives a localization context $\Lambda$, which lists the failing repair tests and the suspicious instructions with their evidence, and returns repair hypotheses, each naming a failure mechanism, the failing tests it explains, and the instructions it affects. $\Propose(h,\Gamma)$ receives a hypothesis $h$ and an edit context $\Gamma$, which offers the edits allowed at each affected instruction \chong{together with the context needed to choose among them}.
\chong{It returns a patch $\Delta$ and the repair tests on which $\Delta$ is to be checked.
These are the source failures $F_\Delta$ that it should fix, control tests} $C_\Delta$ that pass and should not break, and boundary tests $B_\Delta$ at the edge of the intended change.
\end{definition}

These procedures are nondeterministic, and \method assumes nothing about which result a call returns.
We write, for example, $r\in\Propose(h,\Gamma)$ when $r$ is a possible result of that call.
The deterministic \emph{host} builds every argument from \skir, the dependence graphs, and the recorded traces, and checks every result before using it.
It discards addresses that do not resolve, drops hypotheses whose tests are not current failures or whose instructions were not presented, type-checks patches, and validates named tests \chong{against the repair \chong{tests}.
We write $H$ for the hypotheses that \chong{\Diagnose returns and that pass the checks on their tests and instructions}.}

\begin{algorithm}[!t]
\caption{One repair iteration of \method. Lines marked $E$ and $O$ call the execution agent and the repair model, and all other lines are deterministic. \chong{Symbols are defined with their steps.}}
\label{alg:repair}
\begin{algorithmic}[1]
\REQUIRE current skill $S_t$ with its recorded validation outcomes, repair tests $D_{\mathrm{rep}}$, validation tests $D_{\mathrm{val}}$, budget $K$
\STATE $(I_t,M_t)\gets\mathcal{C}(S_t)$; derive \skast, the CFG, and the PDG
\STATE record the trace $\tau_i$ of $\Execute(S_t,x_i)$ for each $i\in D_{\mathrm{rep}}$ \hfill $E$
\STATE \chong{$H\gets$ the checked hypotheses of $\Diagnose(\Lambda_t)$, where $\Lambda_t$ holds the retained suspicious instructions in} $\bigcup_{i\in F(S_t)}\mathit{cov}(\tau_i)$ \hfill $O$
\FORALL{$h\in H$}
\STATE $(\Delta_h,F_{\Delta_h},C_{\Delta_h},B_{\Delta_h})\gets\Propose(h,\Gamma_t(h))$ \hfill $O$
\STATE reject $\Delta_h$ if it is ill-typed
\STATE $S'_h\gets\operatorname{Materialize}(\operatorname{ApplyMutation}(I_t,\Delta_h),M_t)$
\STATE reject $S'_h$ if its recompiled \skir is not well formed; record round-trip observations
\STATE $\Execute(S'_h,x_i)$ for each $i\in L(S'_h)$; compare the outcomes with those of $S_t$ \hfill $E$
\ENDFOR
\STATE $\mathit{Top}\gets$ at most $K$ highest-ranked patched skills that fix a source failure
\STATE $\Execute(S',x_i)$ for each $S'\in\mathit{Top}$ and $i\in D_{\mathrm{val}}$ \hfill $E$
\RETURN an $S'\in\mathit{Top}$ that satisfies the acceptance rule, with its validation outcomes, or else $S_t$
\end{algorithmic}
\end{algorithm}

\subsection{\skir: A Partial Program View of Skills} \label{sec:representation}

To localize a fault and relate it to other instructions, \method needs a skill's operations, conditions, and data references as explicit program elements, yet the execution agent keeps reading the prose.
\method therefore builds a partial program view of the skill.
A deterministic compiler divides the document into instructions and applies a fixed set of \emph{recognizers} that map parts of an instruction to program elements.
\chong{It} keeps all unrecognized content in the instruction's text and records the source span of every instruction and element, so that a change made to an element can be written back to the text.
Like an island grammar~\cite{moonen2001island}, the compiler interprets only the constructs that its recognizers accept, and the rest of the method depends only on the recognized elements and their source spans, not on which textual constructs are recognized.

\parhead{Definitions} A skill document $S$ is a string, and a span $[\alpha,\beta)$ is a range of its character offsets, which denotes the substring $S[\alpha,\beta)$.
\chong{Two spans $[\alpha,\beta)$ and $[\gamma,\eta)$ are disjoint if $\beta\le\gamma$ or $\eta\le\alpha$.}
Appendix~\ref{app:formal} gives the definitions of this subsection with all components spelled out.

\begin{definition}[\skir] A \skir $I$ consists of a finite set $N$ of instructions and a \chong{finite set $P$ of program elements, which are operations, branch conditions, and the merge points} that close the conditions.
Each element $p\in P$ is owned by the instruction $\mathit{own}(p)\in N$ from which it was extracted, and each operation records the data items that it defines and uses.
Each instruction keeps its text in full, including the content that no element represents, together with its section path and its \emph{attributes}, computed from its text alone, such as its modality \chong{(\texttt{MUST}, \texttt{SHOULD}, \texttt{MAY}, or \texttt{NEVER})} and trigger\chong{, the condition that opens a conditional instruction}.
\chong{Its status is} \textsf{partial} if its elements represent only some of its requirements.
Each instruction also has a canonical identifier, stable under offset shifts.
\end{definition}

\begin{definition}[Source map] \label{def:sourcemap} A source map $M$ from $I$ to $S$ assigns a span $M(n)$ of $S$ to each instruction $n\in N$, and a span $M(p)$, its \emph{source anchor}, to each element $p$ that the compiler anchors in the text, such that (i)~the text of $n$ is obtained from $S[M(n)]$ by removing Markdown syntax such as list markers and line breaks; (ii)~$M(n)$ and $M(n')$ are disjoint for distinct instructions $n$ and $n'$; and (iii)~$M(p)$ lies inside $M(\mathit{own}(p))$.
\end{definition}

The compiler is a function $\mathcal{C}$ from skill documents to pairs $(I,M)$.
\chong{We call $(I,M)=\mathcal{C}(S)$ the \emph{partial program view} of $S$, in which $I$ exposes the recognized structure and $M$ anchors it in $S$.}

\begin{definition}[Recognizers] A recognizer extracts program elements from instruction text \chong{together with} their source anchors.
An \emph{operation recognizer} yields operations together with the data items that they define and use; a \emph{guard recognizer} identifies a branch condition $b$, whose merge point and governed operations, $\mathit{gov}(b)$, are established by the compiler using instruction boundaries and document order; and a \emph{reference recognizer} identifies the occurrences of data items, which are matched across instructions.
The compiler $\mathcal{C}$ is parameterized by its set of recognizers.
\end{definition}

\parhead{Recognizers in \method} The compiler uses fixed rules and no language-model calls (line~1; Table~\ref{tab:compile-rules} in Appendix~\ref{app:compile}).
\chong{Fixed rules make compilation deterministic, so recompiling a patched skill checks the patch rather than the variation of another LLM call.}
\chong{The rules implement the document conventions of Section~\ref{sec:skill-background}.}
Headings define sections, and list items and paragraphs\chong{, or the sentences into which the parser splits them,} become instructions.
\chong{Action verbs from a fixed lexicon become operations, which define or use the data items that inline code spans name, as \cue{read} defines and \cue{return} uses them.
Instruction-initial or postposed conditionals such as \cue{if} or \cue{provided that} govern the operations of their own instruction, whereas a \cue{When to use:} clause governs those of the next instruction in its section.}
\chong{Under these rules, the value-span instruction of} Section~\ref{sec:why-representation} yields one \texttt{return} operation, while its exception about units stays in its text.

\chong{The view is partial in two senses.
First, like analyses of incomplete code~\cite{dagenais2008partial}, the analyses \chong{see only relations stated in recognized text}.
Second, patches can still target unrecognized content.}

\parhead{Well-formedness} \chong{Every later step looks up instructions, elements, and relations through the view, so the view must resolve every reference that it hands out.}
We call $(I,M)$ \emph{well formed} if every address of the address contract and every data reference resolves to an instruction, element, or def--use relation of $I$, and $M$ satisfies Definition~\ref{def:sourcemap} for the compiled document.
\method checks well-formedness before \skir and its graphs are used and again for every patched skill.

\subsection{Static Dependence Analysis} \label{sec:program-analysis}

\chong{A fault reported at one instruction often involves others, because the instruction may run only under a condition stated elsewhere or consume a value} that another instruction produces.
\chong{These relations~(L2)} determine what the repair model inspects and which tests a patch may affect.
\method derives them with standard program analyses~\cite{aho2006compilers,ferrante1987pdg} over \skast, the syntax tree of the recognized elements, and lifts the results to instructions, the unit that traces report and patches edit.

\begin{figure}[!t] \centering \includegraphics[width=0.64\linewidth] {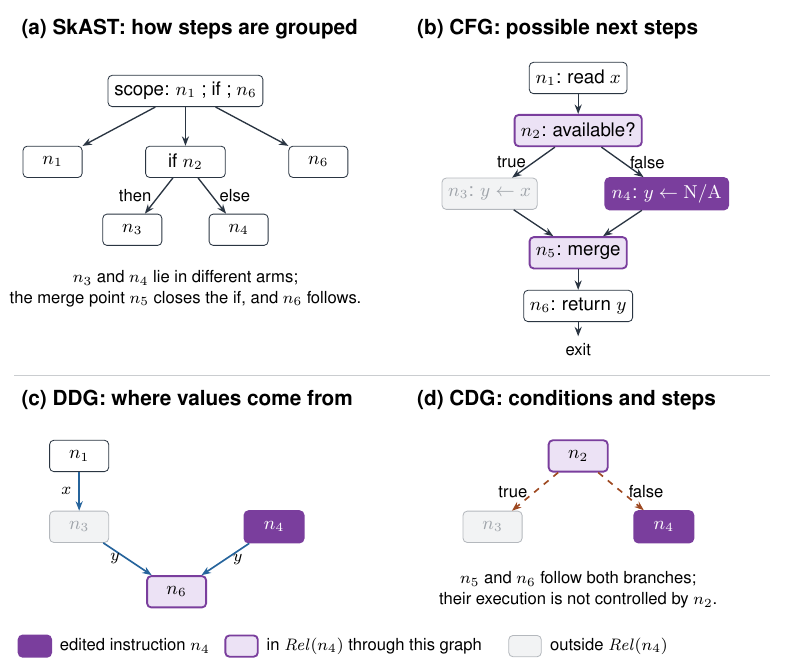} \caption{\skast, CFG, DDG, and CDG of \chong{one schematic} procedure, whose nodes are defined in Appendix~\ref{app:analysis-example}.} \label{fig:analysis-example} \Description{Four panels show a six-node procedure.
The syntax tree contains a read operation, a conditional with two assignments, and a return.
The CFG branches at n2, passes through n3 or n4, merges at n5, and reaches n6.
The DDG connects n1 to n3, and connects n3 and n4 to n6. A patch to n4 is highlighted with its related nodes n2, n5, and n6; n3 is greyed out.
The CDG connects n2 to n3 and n4.}
\end{figure}

\begin{definition}[\skast] \chong{\skast \chong{is the abstract syntax tree of the recognized elements,} a term of the grammar \[ \pi \;::=\; \epsilon \;\mid\; a \;\mid\; \pi_1;\pi_2 \;\mid\; \textsf{if}\ b\ \textsf{then}\ \pi_1\ \textsf{else}\ \pi_2 \;\mid\; \textsf{scope}\ \pi, \] where $a$ ranges over operations and $b$ over branch conditions.
It sequences the operations of each block in source order and places the operations that a branch condition governs in its true or false arm, closed by the condition's merge point, which preserves authored operation order and makes branch scope and rejoining points explicit.
Appendix~\ref{app:formal} gives the four construction rules.}
\end{definition}

\chong{The repair model needs the steps around a suspicious element, and the dependence analyses need an order on the elements, so \method builds} the \chong{control-flow graph (CFG)} over the elements of $P$ and an entry and an exit node \chong{from \skast} by structural induction in the standard way.
\chong{A sequence links its parts.
An \textsf{if} links its condition to the first node of each branch, or to its merge point when the branch adds no nodes, and links the last node of each nonempty branch to its merge point.
A \textsf{scope} and $\epsilon$ add no nodes.}
\chong{The CFG orders the elements but does not say which of them depend on each other, so \method derives the two dependence graphs from it~\cite{ferrante1987pdg}.
The data dependence graph (DDG) links a definition of a data item to each use that it reaches, as computed by reaching-definitions analysis over the CFG~\cite{aho2006compilers}.
The control dependence graph (CDG) has an edge $(b,w)$ if $w$ postdominates one successor of the branch condition $b$ but does not strictly postdominate $b$.
Together, the DDG and CDG form the program dependence graph (PDG).}
We write $A_{\mathrm{cfg}}$, $A_{\mathrm{cdg}}$, and $A_{\mathrm{ddg}}$ for the edge sets.
\chong{The DDG gives the repair model the elements that produce the inputs of a suspicious element and consume its outputs, and the CDG the conditions that decide whether it runs.}
\method uses these graphs only to choose what the repair model sees and which repair tests screen a patch, and \chong{the acceptance rule compares validation outcomes only}.

\method lifts the graphs \chong{from elements} to instructions through a \chong{bounded neighborhood, which consists of the first hop of the backward and forward slices used in program slicing~\cite{weiser1984slicing,horwitz1990slicing} and of adjacent steps in the CFG.
One hop already reaches the steps, producers, consumers, and conditions described above, and it keeps the context of each suspicious instruction small.}

\begin{definition}[Related instructions] For $p\in P$, its immediate structural neighbors are \[ \mathit{Nbr}(p)= \{q\in P\mid (p,q)\ \text{or}\ (q,p)\ \text{is in}\ A_{\mathrm{cfg}}\cup A_{\mathrm{cdg}}\cup A_{\mathrm{ddg}}\}.
\] The instructions related to $n\in N$ are \[ \mathit{Rel}(n)= \{n\}\cup \bigcup_{\substack{p\in P\\ \mathit{own}(p)=n}} \{\mathit{own}(q)\mid q\in\mathit{Nbr}(p)\}.
\] An instruction with no elements in $P$ is related only to itself.
\end{definition}

\parhead{Example} \chong{Figure~\ref{fig:analysis-example} shows all four structures on a schematic procedure in which all three relations matter.
There, a patch to the fallback assignment $n_4$} is related to its governing condition and its consumer through the PDG and to adjacent steps through the CFG, but not to the other branch.

\subsection{Dynamic Analysis and Fault Localization} \label{sec:attribution}

Static dependences describe how instructions may relate, but not which instructions a failing execution involved.
\method obtains this dynamic evidence by running the current skill on the repair tests (line~2).
Localizing a failure to source-linked instructions gives a patch its location \chong{(L1).
Localization starts from coverage, as} spectrum-based fault localization~\cite{jones2002spectrum} does, except that coverage is reported by the interpreter itself rather than measured by instrumentation.

\parhead{Recording traces} \chong{To obtain this coverage, \method asks the execution agent to report which parts of the skill it used.}
\chong{The address contract of \Execute assigns an address to every instruction and every recognized dependency.}
The trace $\tau_i$ of test $i$ under $S$ records the input, the answer, its outcome, and the reported addresses of the recorded call $\Execute(S,x_i)$.
Resolving these addresses against \skir gives the coverage $\mathit{cov}(\tau_i)\subseteq N$.
An instruction address contributes that instruction, and a dependency address contributes the instructions that own its producer and consumer.
\chong{Like the graphs, these reports only choose what the repair model sees and which tests screen a patch.}

\parhead{Fault localization} \chong{The instructions covered by failing tests, $\bigcup_{i\in F(S)}\mathit{cov}(\tau_i)$, are the suspicious locations.
Because a failing trace also covers instructions that played no part in the failure, \method grades each suspicious instruction by the strength of its evidence \chong{and retains those} whose level reaches a fixed threshold.
\chong{The evidence comprises reported uses and \emph{graph witnesses}, which connect a failure to a candidate source instruction through a chain of execution evidence and graph relations checked by the host.
Appendix~\ref{app:localization} gives the other kinds of evidence, the levels, and the thresholds.}
The localization context $\Lambda_t$ contains the failing tests and, for each retained instruction, its text, its one-hop CFG, CDG, and DDG relations, and the repair tests whose traces cover it, passing ones included, so that the repair model can see what a patch should preserve.}

\parhead{Repair hypotheses} \chong{A patch should address a failure mechanism rather than a single failing test, so \Diagnose groups the failures in $\Lambda_t$ into repair hypotheses (line~3).}
The repair model, not the host, decides which failures share a mechanism\chong{, since this judgment requires reading the wrong answers}.
\chong{So that every edit has a location the host can check, the host lists, for each instruction that a hypothesis affects, its \emph{repair targets}, which are the instruction itself,} its recognized elements and dependencies, and the insertion anchor after it\chong{, that is, the position immediately after its source span.
Each target admits at least one repair operator.
The condition that governs the value-span instruction belongs to a separate clause, so the targets of that instruction are the instruction itself and the anchor after it.}
A failure may yield several hypotheses, each leading to a separate patch, and a patch may coordinate edits to several interacting instructions.

\subsection{Patch Generation} \label{sec:repair}

\chong{To be checked before any test runs, a change must name its location and its kind (L1), which free-form rewriting does not require.}
\method therefore requires every change to be a typed \chong{mutation} at a location resolved from \skir and its source map, so that every patch has a location in \skir and in the source text, can be checked before any test runs, and can be compared with earlier patches to the same element.
As in GenProg, which deletes, inserts, or swaps statements~\cite{legoues2012genprog}, the \chong{mutations} come from a fixed set of repair operators, each typed by the kind of target it applies to and by the content it requires.
Let $I_t$ and $M_t$ be the \skir and source map of the current skill $S_t$.

\begin{definition}[Patch] \chong{Each target of a patch is of one of four kinds.
It is an instruction} ($\textsf{Instr}$, whose text or attributes are edited), a branch condition ($\textsf{Cond}$), a dependency ($\textsf{Dep}$, an existing def--use relation or a pair of endpoints for adding one), or an insertion anchor ($\textsf{Anchor}$, the position immediately after an instruction's source span).
Each \emph{repair operator} $o$ has a signature $\kappa_o\Rightarrow\sigma_o$ and a precondition $\mathit{pre}_o$.
It applies to targets of kind $\kappa_o$ whose current state in $I_t$ satisfies $\mathit{pre}_o$, and its new content must satisfy the schema $\sigma_o$.
A single-edit patch is a typed mutation $\Delta=(e,o,c)$ with a target $e:\kappa_o$ resolved from $(I_t,M_t)$ and content $c:\sigma_o$.
For example, the operator that rewrites an instruction has the signature $\textsf{Instr}\Rightarrow\textsf{Text}$ and replaces the instruction's text.
Applying a patch yields $I'=\operatorname{ApplyMutation}(I_t,\Delta)$.
A coordinated patch contains several mutations \chong{whose edits rewrite, delete, or insert after distinct instructions}, all constructed against $I_t$.
\end{definition}

\chong{\chong{The repair model does not edit the skill directly but chooses among the typed \chong{mutations}} that the host offers.}
For each hypothesis $h$, the host builds the edit context $\Gamma_t(h)$, which offers at each target $e$ the legal operators $\mathit{Opt}_t(e)=\{o\mid e:\kappa_o\wedge\mathit{pre}_o(e,I_t)\}$.
For example, a dependency can be removed only if it exists. $\Propose(h,\Gamma_t(h))$ chooses a target, an operator, and new content (line~5), and the host type-checks the choice against $\mathit{Opt}_t(e)$, the current state of the target, and the operator's schema \chong{and rejects} ill-typed patches before they reach the document (line~6).
\chong{The patches of an iteration are constructed and checked independently.}
In \method, the operators delete, insert, or rewrite instructions and edit their conditions, attributes, and dependencies (Table~\ref{tab:operations} and Figure~\ref{fig:patch-prompt} in Appendix~\ref{app:patch}).
The patch in Section~\ref{sec:overview-walkthrough} rewrites the \chong{text of the} value-span instruction.
Because the execution agent reads the document, not \skir, \chong{every} patch is written back to the text (line~7).

\chong{To write a patch back, the host first turns it into edits of source spans.}
The host records the source edits of a patch as $\mathcal{E}_{\Delta}=((\alpha_j,\beta_j,c'_j))_{j=1}^{k}$, where the offsets refer to characters in $S_t$, $c'_j$ includes all host-generated formatting, and an insertion has $\alpha_j=\beta_j$.
The \emph{footprint} $X_\Delta\subseteq N$\chong{, against which untouched text and affected tests are defined, contains the authored target of each edit, that is, the instruction whose text,} attribute, or condition is rewritten, the deleted instruction, the instruction that states the relation for a relation edit, or \chong{the instruction after which an insertion is placed}.

\begin{definition}[Materialization]
For each mutation, the host renders replacement Markdown $c'$ from
the content $c$ \chong{and the current text of the target. A text edit is formatted as the instruction's Markdown, an attribute or condition edit changes only the corresponding expression in the original text, and a relation edit rewrites the text of the instruction that states the relation. The host places $c'$ at a source span $[\alpha,\beta)$, which is the span of its authored target for a rewriting or deletion and the empty span} at the end of its anchor's span
for an insertion. \chong{Materialization applies the edits from the end of the document backward, so that each edit uses its original offsets. For a single span,}
\begin{equation}
S'=\operatorname{Materialize}(I',M_t)
=S_t[0,\alpha)\cdot c'\cdot S_t[\beta,|S_t|).
\label{eq:materialization}
\end{equation}
\end{definition}

\begin{lemma}[Edit locality] \label{lem:locality} If the nonempty intervals in $\mathcal{E}_\Delta$ are pairwise disjoint and no insertion position lies strictly inside one of them, then every fragment \chong{$[\gamma,\eta)$ of $S_t$ such that each edit ends at or before $\gamma$ or starts at or after $\eta$} occurs unchanged in $S'$, shifted by the net length change of earlier edits.
\end{lemma} Appendix~\ref{app:roundtrip} states the lemma with explicit offsets, proves it, and shows when it also preserves the compiled attributes of untargeted instructions.

\parhead{Checking the round trip} The patch was made on \skir but takes effect in the text, so \method recompiles every patched skill before running any test (line~8).
It enforces well-formedness and \chong{records two round-trip observations.
\emph{Target recovery} is whether the recompiled view contains the intended edit, and \emph{stability} is whether every instruction that the patch neither rewrites nor removes keeps} its canonical identifier, text, and compiled attributes.
\chong{Both are recorded rather than enforced, because the agent reads the text, which Edit locality already preserves outside the edited spans, and the tests judge the effect of the edited text.}
With $\mathcal{C}$ as the \emph{get} and materialization as the \emph{put} direction of a bidirectional transformation~\cite{foster2007lenses}, target recovery is a syntactic analogue of the lens law that an update made through the view is visible in the recomputed view.

\subsection{Patch Screening and Validation} \label{sec:selection}

\chong{A patch that fixes its source failures may break other tests that these failures do not reveal, including tests that never reach the edited instruction (L3).}
\chong{For programs,} regression test selection reruns the \emph{modification-traversing} tests, whose executions reach changed code~\cite{rothermel1997safe}, or the tests whose behavior a change may affect through dependences~\cite{binkley1997regression}; change impact analysis derives an \emph{impact set} of affected entities from the coverage of executions that traverse a change~\cite{orso2003leveraging}; and program repair applies such selection to candidate patches~\cite{lou2024rtsapr}.
\chong{As in that impact analysis, \method combines a static impact set with dynamic coverage} (line~9).
For $X\subseteq N$, let $T(X)=\{i\in D_{\mathrm{rep}}\mid\mathit{cov}(\tau_i)\cap X\neq\varnothing\}$ be the repair tests whose traces cover part of $X$.

\begin{definition}[Selected repair tests]
For a patch $\Delta$ with footprint $X_\Delta$, let
\[
Q_\Delta=
\left(\bigcup_{n\in X_\Delta}\mathit{Rel}(n)\right)
\setminus X_\Delta
\]
contain the one-hop CFG, CDG, and DDG neighbors \chong{of the instructions in $X_\Delta$}.
We call $X_\Delta\cup Q_\Delta$ the one-hop \emph{impact set} of $\Delta$.
Let $F_\Delta$, $C_\Delta$, and $B_\Delta$ be the source failures, control tests, and \chong{boundary tests} that \Propose returns with
$\Delta$, as validated by the host.
The selected repair tests of the patched skill $S'$ \chong{obtained from $\Delta$} are
\begin{equation}
L(S')=
F_\Delta\;\cup\;C_\Delta\;\cup\;B_\Delta
\;\cup\;T(X_\Delta\cup Q_\Delta).
\label{eq:related-training-tasks}
\end{equation}
\end{definition}

\chong{Beyond the tests that \Propose names,} $T(X_\Delta)$ holds the repair tests that traverse the change according to the reported coverage, and $T(Q_\Delta)$ adds tests that traverse the rest of the impact set, including tests that do not cover an edited instruction.
Section~\ref{sec:overview-walkthrough} illustrates them for one patch.

\parhead{Ranking and acceptance} \chong{Screening requires a patch to fix at least one of the failures that motivated it and prefers patches that break no previously passing test, and validation then decides.
Running $S'$ on $L(S')$ and comparing its outcomes with those of $S_t$ recorded in line~2 classifies the selected tests in the manner of the fail-to-pass and pass-to-pass tests of SWE-bench~\cite{jimenez2024swebench}.
\emph{Fail-to-pass} tests are those that $S_t$ fails and $S'$ passes, and \emph{regressions} are those that $S_t$ passes and $S'$ fails. $R_{\mathrm{target}}(S')$ counts the fail-to-pass tests among the source failures $F_\Delta$.
\chong{Screening discards the patched skills with $R_{\mathrm{target}}(S')=0$ and ranks the rest by a lexicographic \emph{fitness}, as GenProg ranks variants by a weighted count of the tests they pass~\cite{legoues2012genprog}.
The fitness puts patches without regressions first and then those whose fail-to-pass tests outnumber their regressions, with the groups and ties defined in Appendix~\ref{app:selection}.}
\juan{Computing this fitness on selected repair tests follows repair tools that reduce evaluation cost through test subsets~\cite{fast2010designing}.}
At most $K$ proceed to validation (line~11).
\method evaluates the shortlisted patched skills on the frozen validation suite (line~12) and accepts the one with the highest validation pass rate if $J_{D_{\mathrm{val}}}(S')>J_{D_{\mathrm{val}}}(S_t)$, and otherwise keeps $S_t$ (line~13).
A regression observed in screening therefore does not by itself reject a patch, \chong{since the selected tests are only a subset and the separate validation suite weighs gains against regressions.}
It ranks the patch behind those without regressions, possibly below the top-$K$ cut.}

\parhead{\chong{Recorded outcomes}} \chong{Because answers vary across requests, screening compares a patched skill with the recorded outcomes of $S_t$, and acceptance adopts the accepted skill's recorded validation outcomes as the next baseline, so each decision is reproducible from the logs.}

\parhead{What the selection assumes} \chong{Running only the selected tests is justified if the tests left out \chong{cannot change their outcomes.}}
A test selection is \emph{safe} if it contains every test whose outcome the change can affect~\cite{rothermel1997safe}.
\chong{For deterministic programs under controlled regression testing~\cite{rothermel1997safe},} \chong{safety follows from the semantics, since a test whose execution} does not reach changed code behaves as before.
A skill has no such semantics, because its interpreter is an LLM, so safety can hold only relative to an assumption about the execution agent.
We state that assumption\chong{,} in the manner of analyses that name the behaviors they do not model~\cite{livshits2015soundiness}.
For a repair test $i$, let $\mathit{Ctx}_t(i)=\bigcup_{n\in\mathit{cov}(\tau_i)}\mathit{Rel}(n)$ hold the instructions that its trace covers together with their related instructions, and let $\mathit{Out}_S(i)$\chong{, unlike recorded outcomes,} be the distribution of the outcome, pass or fail, of $\Execute(S,x_i)$.

\begin{definition}[Interpretation locality] \label{def:locality} The execution agent is \emph{one-hop local} for a well-typed patch $\Delta$ at $S_t$ if every repair test $i$ with $X_\Delta\cap\mathit{Ctx}_t(i)=\varnothing$ satisfies $\mathit{Out}_{S'}(i)=\mathit{Out}_{S_t}(i)$, where $S'$ is the patched skill.
It is \emph{coverage local} for $\Delta$ at $S_t$ if the same holds for every $i$ with $X_\Delta\cap\mathit{cov}(\tau_i)=\varnothing$.
\end{definition}

Both conditions have the form of noninterference~\cite{goguen1982security}.
\chong{Intuitively, one-hop locality says that an edit can change the outcome of a test only through the instructions that the test uses or their immediate neighbors in the CFG, CDG, or DDG.}
\chong{Each states that a patch whose footprint lies outside a test's neighborhood does not interfere with the test's outcome.}
Because $n\in\mathit{Rel}(n)$, we have $\mathit{cov}(\tau_i)\subseteq\mathit{Ctx}_t(i)$, so coverage locality implies one-hop locality.
\chong{Coverage locality thus requires unchanged outcomes for more tests, including those that cover a related instruction but none that the patch edits.}

\begin{proposition}[Relative safety of test selection] \label{prop:selection} Let $\Delta$ be a well-typed patch of $S_t$ with patched skill $S'$. (i)~If the execution agent is one-hop local for $\Delta$ at $S_t$, then $\mathit{Out}_{S'}(i)=\mathit{Out}_{S_t}(i)$ for every repair test $i\notin L(S')$. (ii)~The coverage-only selection $L_0(S')=F_\Delta\cup C_\Delta\cup B_\Delta\cup T(X_\Delta)$ has the property of~(i) if the execution agent is coverage local for $\Delta$, but not, in general, under one-hop locality.
\end{proposition} \chong{The proof, given in Appendix~\ref{app:selection}, rests on the symmetry of $\mathit{Rel}$.}

Proposition~\ref{prop:selection} states what the \chong{control-flow and dependence relations} contribute.
They weaken the locality that safe selection requires, from the instructions a test covers to their one-hop neighborhood, which holds the instructions \chong{linked to them by a dependence or an adjacent CFG step}.

\subsection{Properties of an Iteration} \label{sec:guarantees}

\chong{Because the execution agent and the repair model are reached only through calls whose results the host checks, three properties of every iteration hold by construction, whatever the LLMs return.
The next skill compiles to a well-formed partial program view.
An accepted patch leaves the source span of every instruction it does not target unchanged, since the host writes each mutation back only to the spans of its targets, which are disjoint from the spans of all other instructions (Lemma~\ref{lem:locality}).
The recorded validation pass rate never decreases and rises with every accepted patch, so the number of accepted patches is bounded by the number of validation tests that the initial skill fails.
Appendix~\ref{app:selection} states them as a theorem and proves it.}

\chong{These properties and Proposition~\ref{prop:selection} answer the two difficulties of the introduction.
An edit made on incomplete structure lands in the authored text without disturbing the rest, and although the interpreter is an LLM, the choice of screening tests is safe relative to a stated locality assumption, while acceptance rests on running the skill.}

\section{Experiments} \label{sec:experiments}

We evaluate \method on three research questions \chong{and close with a case study of one repair}.

\begin{itemize}

\item \textbf{RQ1: Skill Optimization Effectiveness.}
Can \method improve task performance
compared with the initial skill and
existing skill optimization methods,
Trace2Skill, SkillBoost, and SkillOpt, across different benchmarks
and execution models?

\item \textbf{RQ2: Contribution of \skir-Based Repair \chong{and Screening}.}
What is the combined contribution of \skir-based repair\chong{~(L1, L2)} and the screening of \method\chong{~(L3)}, relative to constrained text editing screened only on
failing repair tests?

\item \textbf{RQ3: Repair Mechanism Analysis.}
\chong{Does \skir-based localization tie repair targets to existing instructions and does write-back recover every well-typed patch in the text (L1), and does program-analysis-guided test selection add screening tests that coverage of the edited instruction misses (L3)?}
\end{itemize}

\subsection{Experimental Setup} \label{sec:exp-setup}

\parhead{Benchmarks} \chong{For DocVQA~\cite{mathew2021docvqa}, we use 534 questions from the official validation split, divided into 107 repair tests, 53 validation tests, and 374 held-out test questions.
For LiveMathematicianBench~\cite{he2026livemathematicianbench} (LiveMath), we use a fixed subset of 177 questions, with 35 repair tests, 17 validation tests, and 125 held-out test questions.
All compared methods use the same partitions\chong{, which the sample identifiers and preparation scripts in the artifact define}.
Both benchmarks pose single-turn questions with reference answers, so answers are scored by rule, without an LLM judge.}

\parhead{\chong{Initial skills}} \juan{\chong{All skill-based methods start from the same benchmark-specific initial skills, released with SkillOpt~\cite{skillopt2026}.}
The instruction labels, rule titles, and applicability clauses \chong{give compilation and repair addressable instruction units and explicit contexts.} 
\chong{Within the DocVQA initial skill, the value-span instruction of} Section~\ref{sec:why-representation} retains the wording of SkillOpt's released GPT-5.5 reference skill, with its condition and action represented separately.}

\parhead{Models and metrics} \chong{GPT-6.1 Sol with medium reasoning effort is the repair model in all experiments, and GPT-5.6 Luna, GPT-6 Sol, and Gemini 3.6 Flash are the execution models on both benchmarks.
We report the pass rate $J_{D_{\mathrm{test}}}$ on the held-out tests (Section~\ref{sec:problem}) as the task success rate.
DocVQA counts an answer as correct when its average normalized Levenshtein similarity (ANLS)~\cite{biten2019stvqa} is at least $0.999$, and LiveMath uses SkillOpt's multiple-choice scoring rule.}

\parhead{Baselines}\label{sec:baselines} \chong{We compare \method with the Initial Skill, which measures the improvement brought by optimization, with No Skill, which runs each execution model without a skill document, and with three methods that revise natural-language skills from execution feedback.
Trace2Skill~\cite{trace2skill2026} merges patches distilled from individual trajectories, SkillBoost~\cite{lin2026skillboost} bounds regressions on a fixed task set, and SkillOpt~\cite{skillopt2026} applies bounded patches accepted on a \chong{split held out from patch generation}.
All methods use the same execution models and scoring rules within each setting.
Trace2Skill performs one trajectory-analysis and patch-merging pass and selects between its initial and evolved skill by validation performance.
SkillOpt runs four epochs on DocVQA and three on LiveMath \chong{and is evaluated on the held-out tests at the checkpoint with the highest validation success rate}.
\juan{SkillBoost uses budgeted rounds with Best-of-4 generation and Top-2 evaluation on the full repair-test set, and selects the checkpoint with the highest validation success rate}.
\method runs three repair \chong{iterations} with the selection procedure of Section~\ref{sec:selection}.
\juan{All skill-optimization methods use GPT-6.1 Sol with medium reasoning effort, \chong{the repair model of \method, to revise their skills}.}
\juan{The execution interfaces preserve each benchmark's task inputs and required answer format.
\method additionally supplies an address contract and \chong{requests a usage report (Section~\ref{sec:attribution}).}}}

\parhead{Repair protocol} \chong{For each benchmark and execution model, a repair run has three repair iterations, at most $K=4$ patches per iteration proceed to validation, and the final selected skill is evaluated on the held-out tests.
Test scores of earlier method versions were recorded during development.}

\subsection{Skill Optimization Effectiveness} \label{sec:effectiveness}

\chong{Table~\ref{tab:main-results} shows that \method achieves the highest \juan{test success rate} in all six settings (RQ1).
\chong{\method} exceeds SkillOpt by \juan{3.74--4.54} percentage points on DocVQA and 2.40--4.00 on LiveMath, Trace2Skill by 1.87--4.54 and 5.60--10.40, and SkillBoost by 1.60--5.88 and 3.20--13.60.}

\begin{table}[!htbp]
\centering
\caption{Task success rate (\%) on DocVQA and LiveMath. Each value is a success rate on the held-out tests.
The best result in each row is in bold.
Percentages are truncated, not rounded, to two decimals.
\juan{Differences in the text are computed from untruncated rates and then truncated.}}
\label{tab:main-results}
\papertablestyle
\setlength{\tabcolsep}{4pt}
\begin{tabular}{@{}cccccccc@{}}
\toprule
Benchmark & \chong{Execution model} & No Skill & Initial Skill
& Trace2Skill & SkillBoost & SkillOpt & \method \\
\midrule

\multirow{3}{*}{DocVQA}
& GPT-5.6 Luna
& 55.61 & 71.92 & 82.88 & 81.55 & 82.35 & \textbf{86.63} \\

& GPT-6 Sol
& 62.56 & 77.54 & 86.09 & 84.75 & 86.09 & \textbf{90.64} \\

& Gemini 3.6 Flash
& 90.64 & 91.17 & 93.58 & 93.85 & 91.71 & \textbf{95.45} \\

\midrule

\multirow{3}{*}{LiveMath}
& GPT-5.6 Luna
& 31.20 & 36.80 & 36.00 & 32.80 & 44.00 & \textbf{46.40} \\

& GPT-6 Sol
& 48.00 & 54.40 & 55.20 & 55.20 & 58.40 & \textbf{62.40} \\

& Gemini 3.6 Flash
& 32.80 & 32.80 & 32.00 & 34.40 & 34.40 & \textbf{37.60} \\

\bottomrule
\end{tabular}
\end{table}

\parhead{\chong{Strongest baseline}} \chong{No baseline is the strongest in every setting.
Trace2Skill leads the baselines for GPT-5.6 Luna on DocVQA, SkillBoost for Gemini 3.6 Flash on DocVQA, and SkillOpt for GPT-5.6 Luna and GPT-6 Sol on LiveMath, and two baselines tie in each of the other two settings.
\method exceeds the strongest baseline of every setting, by 1.60 to 4.54 points.
Measured from the initial skill, its gain is 1.3 to 3.0 times the gain of the strongest baseline in each setting.}

\parhead{\chong{Gains over the initial skill}} \chong{The Initial Skill helps much on DocVQA for GPT-5.6 Luna and GPT-6 Sol, little for Gemini 3.6 Flash, less on LiveMath, and not at all for Gemini 3.6 Flash on LiveMath.
On DocVQA, \method removes about half of the held-out errors of the initial skill with every execution model (48--58\%), even for Gemini 3.6 Flash, whose initial skill already answers 91.17\% of the questions correctly, while the strongest baseline of each setting removes 30--39\%.
This share stays near one half although the success rate of the initial skill ranges from 71.92\% to 91.17\%.
On LiveMath, \method removes 7--18\% of the errors of the initial skill, and the strongest baseline 2--11\%.
\method thus improves on the initial skill in every setting, which not every optimizer does.
Trace2Skill and SkillBoost end below the initial skill on LiveMath with GPT-5.6 Luna (36.00 and 32.80 versus 36.80), and Trace2Skill also with Gemini 3.6 Flash (32.00 versus 32.80), whereas \method and SkillOpt improve on it in all six settings.}

\parhead{\chong{Skill versus execution model}} \chong{On DocVQA, the repaired skill can outweigh a stronger execution model.
With the skill repaired by \method, GPT-5.6 Luna answers 86.63\% of held-out questions, more than GPT-6 Sol with the initial skill (77.54\%) or with any baseline's final skill (at most 86.09\%).}

\parhead{\chong{Size of the change}} \chong{Each of the three repair iterations accepts at most one patch, so every final skill of \method results from at most three accepted patches, and by Section~\ref{sec:guarantees}, each of them leaves the source span of every instruction it does not target unchanged.
The gains above therefore come from a few edits confined to the instructions they target.}

\parhead{\chong{Answer to RQ1}} \chong{\method achieves the highest success rate in every setting, exceeds the strongest baseline of each setting by 1.60 to 4.54 points, removes about half of the initial skill's errors on DocVQA, and never ends below the initial skill, with at most three localized patches per skill.}

\subsection{Contribution of \skir-Based Repair \chong{and Screening}} \label{sec:ablation}

\chong{To measure the combined contribution of \skir-based repair and the screening of \method (RQ2), we compare the full \method approach with a variant that replaces \skir-based repair with constrained text editing and narrows screening.}
The comparison keeps the initial skill, repair \chong{tests}, execution models, dataset splits, repair iterations, and the validation-based acceptance rule fixed.
The w/o \skir variant differs from \method in two respects.
First, it removes \skir, \skast, CFG, CDG, DDG, source-map links, and typed \skir mutations, \chong{which it replaces with constrained text editing}\juan{, with the host limiting each patch to at most 120 line additions and deletions in total}\chong{, while retaining execution traces}.
The repair model still receives the complete repair-test results of the current skill.
Second, it screens each \chong{patched skill} only on the failing repair tests of the current skill, $F(S_t)$, instead of on $L(S')$ (Equation~\ref{eq:related-training-tasks}).
It thus omits the \chong{control tests} $C_\Delta$, the boundary tests $B_\Delta$, and the coverage-based tests $T(X_\Delta\cup Q_\Delta)$, so its screening observes no regressions\juan{, and only validation can reject a patch that breaks passing tests}.
Ranking and validation are otherwise unchanged.

\begin{table}[!htbp]
\centering
\caption{Effect of \skir-based repair \chong{and its screening}
on DocVQA and LiveMath. Results are
test success rates (\%). Higher is better. The w/o \skir variant
replaces \skir-based repair with constrained text editing
and screens candidates only on failing repair tests.}
\label{tab:ir-ablation}
\papertablestyle
\begin{tabular}{@{}cccc@{}}
\toprule
Benchmark & \chong{Execution model}
& Full \method
& w/o \skir \\
\midrule

\multirow{2}{*}{DocVQA}
& GPT-5.6 Luna
& 86.63
& 84.22 \\

& GPT-6 Sol
& 90.64
& 87.97 \\

\midrule

\multirow{2}{*}{LiveMath}
& GPT-5.6 Luna
& 46.40
& 44.80 \\

& GPT-6 Sol
& 62.40
& 59.20 \\

\bottomrule
\end{tabular}
\end{table}

\chong{Replacing \skir-based repair and its screening with the w/o \skir variant reduces \juan{test success rates} in all four evaluated settings (Table~\ref{tab:ir-ablation}), by \juan{2.40} and 2.67 percentage points on DocVQA and 1.60 and 3.20 on LiveMath for GPT-5.6 Luna and GPT-6 Sol.
\chong{The drop is larger for GPT-6 Sol than for GPT-5.6 Luna on both benchmarks.}
\juan{The broader screening gives the ranking results from control tests, boundary tests, and structurally related repair tests, so that patches that fix their source failures while keeping the selected passing tests are preferred before validation.}
}

\chong{At the question level, \juan{in the GPT-6 Sol DocVQA test evaluation,} Full \method answers \juan{17 questions} that w/o \skir misses, and w/o \skir answers 7 that Full \method misses, \juan{a net gain} of 10 of 374 questions.
\juan{Full \method achieves this held-out gain with a validation score of 50 of 53, compared with 51 of 53 for w/o \skir.}}

\parhead{\chong{Answer to RQ2}} \chong{Replacing \skir-based repair and its screening with constrained text editing lowers success rates in all four settings, so together they contribute to the gains of \method.}

\subsection{Repair Mechanism Analysis} \label{sec:repair-analysis}

To examine how the mechanisms behind L1 and L3 operate (RQ3), we analyze two recorded DocVQA runs of the full \method, one with GPT-5.6 Luna and one with GPT-6 Sol.
Table~\ref{tab:repair-mechanisms} summarizes the four stages of repair that the analysis follows.

\begin{table}[t]
\centering
\caption{Repair mechanism analysis from two DocVQA runs of the
full \method, one with GPT-5.6 Luna and one with GPT-6 Sol.
\chong{Counts pool the three iterations of both runs, and each row is one stage of repair.}}
\label{tab:repair-mechanisms}
\papertablestyle
\setlength{\tabcolsep}{5pt}
\renewcommand{\tabularxcolumn}[1]{m{#1}}
\begin{tabularx}{\linewidth}{@{}>{\centering\arraybackslash}m{0.20\linewidth}>{\centering\arraybackslash}X>{\centering\arraybackslash}m{0.30\linewidth}@{}}
\toprule
Repair stage
& Metric
& Result \\
\midrule

Localization & \chong{Repair targets that are existing instructions rather than insertion anchors}
& 292/315 (92.70\%) \\

\chong{Write-back} & Well-typed \chong{patches} \chong{passing the round-trip observations}
& 37/37 (100.00\%) \\

\chong{Source-failure fix} & \chong{Patches fixing a source failure}
& 19/37 (51.35\%) \\

\chong{Test selection} & \chong{Patches with tests added through related instructions}
& 23/37 (62.16\%) \\

\bottomrule
\end{tabularx}
\end{table}

\parhead{\chong{Localization and write-back}} \chong{The first two stages give each patch the location and kind of change that L1 requires.
Across six iterations of the two runs, 292 of the 315 recorded repair targets are existing instructions with valid source mappings and 23 are insertion anchors, so most failures localize to an instruction that a patch can rewrite rather than to a gap that needs a new one.
Of the 38 proposals that the repair model produces from these targets, 37 pass type checking, and all 37 are written back as complete skill documents that pass the round-trip observations of Section~\ref{sec:repair}.}

\parhead{\chong{Source-failure fixes}} \chong{Screening first requires a patch to fix a failure that motivated it.
Of the 37 patches, each evaluated on its selected repair tests, 19 (51.35\%) fix at least one of their source failures ($R_{\mathrm{target}}(S')>0$), and screening discards the other 18 before any validation call.
Appendix~\ref{app:archive} compares evaluated and accepted patches in archived trajectories.}

\parhead{\chong{Where repair falls short}} \chong{The discarded patches fail in two ways.
Some leave the target error unchanged.
In the GPT-6 Sol run, a patch that tells the agent to return multiple rows only when they tie still yields the same incorrect two-name answer to a question that asks for the unique maximum.
Others move an answer toward the reference without reaching the ANLS threshold, such as three patches that raise the ANLS of one answer from 0.720 to 0.885.
Fixing a source failure does not guarantee a net gain either.
A GPT-5.6 Luna patch corrects a page-number answer from \cue{2 OF 2} to \cue{2} but recovers three answers and regresses on six others, so screening ranks it below every patch whose fixes outnumber its regressions.}

\parhead{\chong{Program-analysis-guided test selection}} Coverage of the related instructions adds repair tests beyond the source failures, \chong{control tests}, boundary tests, and coverage of the edited instructions~(L3) for 14 of 21 \chong{GPT-5.6 Luna} \chong{patches} and 9 of 16 GPT-6 Sol \chong{patches}, 23 of 37 overall (62.16\%).
\chong{The added tests cover related instructions, and under the one-hop locality assumption of Proposition~\ref{prop:selection}, a patch can affect a test beyond its direct reported coverage only if the test covers such an instruction.
These are exactly the tests that the coverage-only selection $L_0$ of Proposition~\ref{prop:selection} leaves out, so for 23 of the 37 patches, selection through the relations checks tests that coverage alone would not.}
\chong{In the case-study run of Section~\ref{sec:overview-walkthrough}, coverage of a CFG neighbor adds tests} associated with the similar-strings condition, \chong{and the receipt test of Section~\ref{sec:overview-walkthrough} is one of them}.

\parhead{Answer to RQ3} \chong{Localization ties most repair targets to existing instructions that a patch can rewrite, and write-back recovers every well-typed patch before any test runs (L1).
Related instructions add tests for 23 of 37 patches, including one whose regression coverage misses (L3).}

\subsection{\chong{Case Study}} \label{sec:case-study}

\chong{The repair of Section~\ref{sec:overview-walkthrough} lets us follow each limitation through a recorded run.
It comes from a separate historical DocVQA run with GPT-6 Sol as the execution model, and Table~\ref{tab:turnover} compares the final answers of the four methods to the held-out turnover question, all with the same execution model, initial skill, and partitions.}

\begin{table}[!t]
\centering
\caption{\chong{Final answers to the held-out turnover question with GPT-6 Sol (reference \cue{41,809.82~crores}).}}
\label{tab:turnover}
\papertablestyle
\begin{tabular}{@{}lll@{}}
\toprule
Method & Final answer & Assessment \\
\midrule
\method & 41,809.82 crores & Correct \\
SkillOpt & 41,809.82 & Incomplete \\
SkillBoost & 41,809.82 & Incomplete \\
Trace2Skill & 41,809.82 & Incomplete \\
\bottomrule
\end{tabular}
\end{table}

\parhead{\chong{Location and kind of change (L1)}} \chong{The traces of the failing repair tests map to the value-span instruction, and the accepted patch is a single rewrite of that instruction at its mapped span (line~31 of the initial skill).
Recompilation yields a well-formed view, and no other source span changes.}

\parhead{\chong{Related instructions (L2)}} \chong{The request for the patch contains the condition that governs the instruction, linked by a CDG edge from the \cue{When to use} clause at line~29, and the neighboring condition for similar strings at line~35, linked by a CFG edge, and the saved response matches the applied edit.}

\parhead{\chong{Screening (L3)}} \chong{The patch fixes both failing repair tests that motivated it.
Among the named passing tests, table cells keep their concise numeric answers, and monetary fields keep the currency symbols printed with their values.
The receipt test, which enters screening only through the neighboring condition, regresses, so the patch ranks behind regression-free patches.}

\parhead{\chong{Outcome}} \chong{Generated in the first repair iteration, the patch passes more validation tests than the initial skill, is accepted, and remains the final selected skill.
On the held-out turnover question, the revised instruction keeps \cue{crores}, which specifies the scale of the amount and serves as its unit of expression, whereas the final skills of SkillOpt, SkillBoost, and Trace2Skill drop it (Table~\ref{tab:turnover}).
Appendix~\ref{app:d010-example} gives the identifiers of the run, the instructions, and the patch.}

\section{Discussion and Limitations}
\label{sec:threats}

\parhead{Beyond single-turn question answering}
\chong{The partial program view does not depend on the task format.
Skills for multi-step tool use state the same kinds of elements, namely operations, the conditions under which they apply, and the data they pass on, and the address contract can label the instructions that request a tool call.
What changes is the trace, which spans several turns, so coverage and outcomes would be recorded per turn and a test could fail at an intermediate step.}

\parhead{Recognized structure}
\chong{The analyses see only the structure that the fixed recognizers of Section~\ref{sec:representation} accept, so a skill that uses other verbs, refers to an item by different phrases, or describes iteration in prose yields sparser dependence graphs, and the initial skills of our benchmarks may suit these rules better than skills written in other styles.
Richer recognizers, including learned ones, could extend the view, provided that compilation stays deterministic and passes the same checks.}

\parhead{Self-reported coverage and locality}
\chong{Localization and test selection rely on the instruction usage that the execution agent reports, and test selection is safe only for a one-hop local agent (Proposition~\ref{prop:selection}).
The properties of Section~\ref{sec:guarantees} hold whether or not the reports are accurate, but this paper measures neither the reports nor the locality assumption.
Counterfactual checks, such as removing an instruction and rerunning the tests that report using it, could validate coverage, and rerunning all repair tests for sampled patches would show how often locality fails.}

\parhead{One-hop neighborhoods}
\chong{The repair model sees, and screening adds, only the instructions one hop away from a suspicious or edited instruction.
An effect that travels further, such as the influence of the condition that lies two hops from the return in Figure~\ref{fig:analysis-example}, is covered only if the agent's reports or the named tests reach it.
Neighborhoods that grow with the skill, or that follow dependence chains beyond the instructions a test covers, would trade a larger context for wider coverage.}

\parhead{Acceptance on a small validation suite}
\chong{\method accepts a patch when it passes at least one more validation test than the current skill.
With small suites, such as the 53 and 17 validation tests here, and answers that vary across requests, a patch can be accepted by chance~\cite{cawley2010overfitting}.
Requiring a margin or repeated validation runs would make acceptance more robust at a higher cost.}

\section{Related Work}
\label{sec:related-work}

\parhead{Automatic skill optimization}
\chong{Skill optimization continues work that improves LLM behavior from feedback without changing model parameters, such as the verbal reflections of Reflexion~\cite{reflexion2023} and the feedback that TextGrad passes through a computation graph~\cite{textgrad2024}.
Trace-driven methods revise a skill from execution trajectories~\cite{skillgrad2026}, SkillOpt bounds text edits and accepts them only after validation improvement~\cite{skillopt2026}, and Trace2Skill consolidates trajectory-local patches with conflict and format checks~\cite{trace2skill2026}.
Attribution-guided methods first localize a failure to skill sections, fields, or mechanism relations~\cite{reskill2026,lin2026wml}.
These methods express the edit as rewritten text, whereas \method localizes to source-linked instructions, edits them through typed mutations~(L1), and relates them to other instructions through control and data dependences~(L2).
Structure-aware methods are the closest in how they edit.
SEPO applies typed operators to the steps and rules of a prompt outline~\cite{ma2026sepo}, and GraphSkillEvo mutates the step nodes and workflow paths of a graph-structured skill~\cite{sun2026graphskillevo}.
They edit a structured representation of the prompt or skill, whereas \method recovers structure from the prose of an existing skill and writes every edit back to it.
We do not include SEPO, GraphSkillEvo, and GraphSkillAA as baselines, because their editing units are prompt outlines and graph-structured skills rather than shared skill documents.
Library-level methods expand, consolidate, and relate skill collections~\cite{skillcommit2026}, whereas \method revises the instructions within one skill.
To check an edit, SkillBoost bounds regressions on a fixed task set~\cite{lin2026skillboost} and SkillCommit rechecks source tasks~\cite{skillcommit2026}.}
\juan{The closest concurrent approach, GraphSkillAA~\cite{skillaa2026}, applies typed edits to an addressable skill graph and uses a Local Gate to retest affected examples.
\method \chong{instead} grounds both editing and test selection in a source-linked partial program view recovered from the authored skill text.
Its source map confines typed edits to their mapped source spans and preserves the surrounding text, while control-flow and dependence relations derived from the same view select related repair tests, including those outside the edited instruction's direct coverage\chong{~(L3)}.}

\parhead{Prompt and context optimization}
\chong{Prompt optimization searches over instructions for a fixed task~\cite{zhou2023ape,pryzant2023protegi}, DSPy and GEPA optimize or reflectively rewrite the prompts of the modules of an LM program~\cite{khattab2024dspy,agrawal2026gepa}, Trace propagates feedback along a workflow's execution graph~\cite{cheng2024trace}, SAMMO and MPO edit the components of a prompt separately~\cite{schnabel2024sammo,sharma2026mpo}, and ACE revises playbook entries for agent contexts~\cite{zhang2026ace}.
These systems optimize the components of LM programs, whereas \method analyzes the dependences among the instructions of one skill.}

\parhead{Structured skill representations}
\chong{Agent Workflow Memory learns reusable procedures from experience~\cite{wang2024agentworkflow}, SSL organizes skills into usage conditions, steps, actions, and resources~\cite{liang2026ssl}, SkillSmith compiles skills into runtime interfaces with operator contracts~\cite{xu2026skillsmith}, SkVM extracts step dependencies to run independent steps in parallel~\cite{chen2026skvm}, and SkillZip builds section graphs to compress skill libraries~\cite{tan2026skillzip}.
These approaches organize skills for reuse, execution, or compression, whereas \method uses recovered structure to modify an existing skill.}

\parhead{Program analysis and prompts as programs}
\chong{\method applies standard ASTs, control-flow graphs, and program dependence graphs~\cite{aho2006compilers,ferrante1987pdg} to the structure extracted from skills.
Like analyses of incomplete code~\cite{dagenais2008partial} and island grammars~\cite{moonen2001island}, it analyzes only the recognized part of its input, and like origin tracking~\cite{vandeursen1993origin} and lenses~\cite{foster2007lenses}, it carries edits made on a derived structure back to the source text.
Prompt programming languages such as LMQL let authors write prompt structure explicitly~\cite{beurerkellner2023lmql}, whereas \method recovers structure from existing prose, and empirical work shows that prompt programs need tool support~\cite{liang2025promptsprograms}.
PromptPex generates tests from the input specifications and output rules of a prompt~\cite{sharma2025promptpex}, and ContextCov compiles agent instruction files into executable checks~\cite{sharma2026contextcov}.
These works test or enforce what a prompt requires, whereas \method analyzes the instructions themselves to decide where to edit and which tests to rerun.}

\parhead{Automated program repair}
\chong{Generate-and-validate repair generates candidate patches with repair operators and validates them by running tests.
GenProg mutates statements and ranks variants by a test-based fitness~\cite{legoues2012genprog}, and LLM-based repair keeps this structure with an LLM as the generator, from zero-shot patch generation~\cite{xia2022alpharepair} to agents and pipelines that localize faults, generate patches, and run tests~\cite{bouzenia2024repairagent,xia2024agentless}.
A plausible patch, which passes all tests used to develop it~\cite{qi2015test}, can still fail on independent tests~\cite{smith2015overfitting}, so \method separates selection on repair and validation tests from estimation on held-out tests~\cite{cawley2010overfitting}.
Section~\ref{sec:selection} discusses regression test selection and impact analysis~\cite{rothermel1997safe,binkley1997regression,orso2003leveraging,lou2024rtsapr}.
Unlike program repair, \method makes edits on a recovered structure while an LLM reads the text, so the source map carries each edit back to the skill, and running the skill shows its effects.}

\section{Conclusion}
\label{sec:conclusion}

\chong{We cast natural-language skill optimization as generate-and-validate repair over a partial program view. \textbf{\method} compiles a skill into \skir so that fault localization, dependence analysis, and regression test selection decide where to edit, what an edit relates to, and which tests to recheck. It writes every typed mutation back to the mapped span of its target, and the properties of each iteration hold by construction whatever the LLMs return. On DocVQA and LiveMath, \method achieves higher \juan{success rates} than SkillOpt, Trace2Skill, and SkillBoost in all six settings, and an ablation of its \skir-based mechanisms and broader screening lowers success rates in all four evaluated settings. The repair of the turnover example shows the pattern behind these results, a single rewrite at the mapped span of one instruction, informed by the condition that governs it and checked on a test reached only through its neighbor. More broadly, the approach shows how program repair can work on a language with only a partial grammar and an LLM as its interpreter, by editing a recovered view whose edits stay within their spans. Future work will apply \method to skills for multi-step tool use and to a broader range of agent tasks.}

\bibliographystyle{ACM-Reference-Format}
\bibliography{sample-base}

@article{trace2skill2026,
  author        = {Ni, Jingwei and Liu, Yihao and Liu, Xinpeng and Sun, Yutao and Zhou, Mengyu and Cheng, Pengyu and Wang, Dexin and Zhao, Erchao and Jiang, Xiaoxi and Jiang, Guanjun},
  title         = {{Trace2Skill: Distill Trajectory-Local Lessons into Transferable Agent Skills}},
  journal       = {arXiv preprint arXiv:2603.25158},
  year          = {2026},
  doi = {10.48550/arXiv.2603.25158},
  eprint        = {2603.25158},
  archivePrefix = {arXiv},
  primaryClass  = {cs.AI},
  url           = {https://arxiv.org/abs/2603.25158}
}

@article{skillopt2026,
  author        = {Yang, Yifan and Gong, Ziyang and Huang, Weiquan and Yang, Qihao and Zhou, Ziwei and Huang, Zisu and Li, Yan and Gao, Xuemei and Dai, Qi and Liu, Bei and Qiu, Kai and Yang, Yuqing and Chen, Dongdong and Yang, Xue and Luo, Chong},
  title         = {{SkillOpt: Executive Strategy for Self-Evolving Agent Skills}},
  journal       = {arXiv preprint arXiv:2605.23904},
  year          = {2026},
  doi = {10.48550/arXiv.2605.23904},
  eprint        = {2605.23904},
  archivePrefix = {arXiv},
  primaryClass  = {cs.AI},
  url           = {https://arxiv.org/abs/2605.23904}
}

@article{skillgrad2026,
  author        = {Wang, Hanyu and Lan, Yifan and Cao, Bochuan and Lin, Lu and Chen, Jinghui},
  title         = {{SkillGrad: Optimizing Agent Skills Like Gradient Descent}},
  journal       = {arXiv preprint arXiv:2605.27760},
  year          = {2026},
  doi = {10.48550/arXiv.2605.27760},
  eprint        = {2605.27760},
  archivePrefix = {arXiv},
  primaryClass  = {cs.AI},
  url           = {https://arxiv.org/abs/2605.27760}
}

@article{xu2026skillsmith,
  title = {{SkillSmith: Compiling Agent Skills into Boundary-Guided Runtime Interfaces}},
  author = {Xu, Duling and Chen, Zheng and Pan, Zaifeng and Guan, Jiawei and Dong, Dong and Li, Jialin and Pu, Bangzheng},
  journal = {arXiv preprint arXiv:2605.15215},
  year = {2026},
  url = {https://arxiv.org/abs/2605.15215}
}

@article{chen2026skvm,
  title = {{SkVM: Revisiting Language VM for Skills across Heterogenous LLMs and Harnesses}},
  author = {Chen, Le and Feng, Erhu and Xia, Yubin and Chen, Haibo},
  journal = {arXiv preprint arXiv:2604.03088},
  year = {2026},
  url = {https://arxiv.org/abs/2604.03088}
}

@article{liang2026ssl,
  title = {{From Skill Text to Skill Structure: The Scheduling-Structural-Logical Representation for Agent Skills}},
  author = {Liang, Qiliang and Wang, Hansi and Liang, Zhong and Liu, Yang},
  journal = {arXiv preprint arXiv:2604.24026},
  year = {2026},
  doi = {10.48550/arXiv.2604.24026},
  eprint = {2604.24026},
  archivePrefix = {arXiv},
  primaryClass = {cs.CL},
  url = {https://arxiv.org/abs/2604.24026}
}

@inproceedings{xia2022alpharepair,
  title = {{Less Training, More Repairing Please: Revisiting Automated Program Repair via Zero-Shot Learning}},
  author = {Xia, Chunqiu Steven and Zhang, Lingming},
  booktitle = {Proceedings of the 30th ACM Joint European Software Engineering Conference and Symposium on the Foundations of Software Engineering},
  year = {2022},
  pages = {959--971},
  publisher = {Association for Computing Machinery},
  doi = {10.1145/3540250.3549101},
  url = {https://arxiv.org/abs/2207.08281}
}

@article{reskill2026,
  author        = {Mengyi Deng and Xin Li and Duyi Pan and Zilin Wang and Zhiwei Li and Zhijiang Guo and Wei Wang},
  title         = {{RESKILL: Explicit Failure Attribution and Structured Repair for Interactive Language Agents}},
  journal       = {arXiv preprint arXiv:2609.15684},
  year          = {2026},
  eprint        = {2609.15684},
  archivePrefix = {arXiv},
  primaryClass  = {cs.CL},
  url           = {https://arxiv.org/abs/2609.15684}
}

@article{skillcommit2026,
  author        = {Yu He and Weikai Yang},
  title         = {{SkillCommit: Evolving Agent Skills through Behaviorally Validated Scope Expansion}},
  journal       = {arXiv preprint arXiv:2608.15165},
  year          = {2026},
  eprint        = {2608.15165},
  archivePrefix = {arXiv},
  primaryClass  = {cs.AI},
  url           = {https://arxiv.org/abs/2608.15165}
}

@article{skillaa2026,
  author        = {Ziqiao Shang and Ling-Yue Ge and Lan-Zhe Guo},
  title         = {{GraphSkillAA: Attribution-Guided Skill-Graph Updating with Targeted Validation and Rollback}},
  journal       = {arXiv preprint arXiv:2609.20455},
  year          = {2026},
  eprint        = {2609.20455},
  archivePrefix = {arXiv},
  primaryClass  = {cs.AI},
  url           = {https://arxiv.org/abs/2609.20455}
}

@inproceedings{react2023,
  author       = {Shunyu Yao and Jeffrey Zhao and Dian Yu and Nan Du and Izhak Shafran and Karthik R. Narasimhan and Yuan Cao},
  title        = {{ReAct: Synergizing Reasoning and Acting in Language Models}},
  booktitle    = {{International Conference on Learning Representations}},
  year         = {2023},
  eprint       = {2210.03629},
  archivePrefix = {arXiv},
  primaryClass = {cs.CL},
  url          = {https://arxiv.org/abs/2210.03629}
}

@inproceedings{schick2023toolformer,
  author = {Schick, Timo and Dwivedi-Yu, Jane and Dess{\`\i}, Roberto and Raileanu, Roberta and Lomeli, Maria and Hambro, Eric and Zettlemoyer, Luke and Cancedda, Nicola and Scialom, Thomas},
  title = {{Toolformer}: Language Models Can Teach Themselves to Use Tools},
  booktitle = {Advances in Neural Information Processing Systems},
  volume = {36},
  year = {2023},
  url = {https://proceedings.neurips.cc/paper/2023/hash/d842425e4bf79ba039352da0f658a906-Abstract.html}
}

@inproceedings{yang2024sweagent,
  author = {Yang, John and Jimenez, Carlos E. and Wettig, Alexander and Lieret, Kilian and Yao, Shunyu and Narasimhan, Karthik and Press, Ofir},
  title = {{SWE-agent}: Agent-Computer Interfaces Enable Automated Software Engineering},
  booktitle = {Advances in Neural Information Processing Systems},
  volume = {37},
  year = {2024},
  doi = {10.52202/079017-1601},
  url = {https://papers.neurips.cc/paper_files/paper/2024/hash/5a7c947568c1b1328ccc5230172e1e7c-Abstract-Conference.html}
}

@article{wang2024agentsurvey,
  author = {Wang, Lei and Ma, Chen and Feng, Xueyang and Zhang, Zeyu and Yang, Hao and Zhang, Jingsen and Chen, Zhiyuan and Tang, Jiakai and Chen, Xu and Lin, Yankai and Zhao, Wayne Xin and Wei, Zhewei and Wen, Jirong},
  title = {A Survey on Large Language Model Based Autonomous Agents},
  journal = {Frontiers of Computer Science},
  volume = {18},
  number = {6},
  pages = {186345},
  year = {2024},
  doi = {10.1007/s11704-024-40231-1}
}

@inproceedings{reflexion2023,
  author = {Shinn, Noah and Cassano, Federico and Gopinath, Ashwin and Narasimhan, Karthik and Yao, Shunyu},
  title = {{Reflexion}: Language Agents with Verbal Reinforcement Learning},
  booktitle = {Advances in Neural Information Processing Systems},
  volume = {36},
  pages = {8634--8652},
  year = {2023},
  publisher = {Curran Associates, Inc.},
  doi = {10.52202/075280-0377},
  url = {https://proceedings.neurips.cc/paper_files/paper/2023/hash/1b44b878bb782e6954cd888628510e90-Abstract-Conference.html}
}

@article{textgrad2024,
  author = {Yuksekgonul, Mert and Bianchi, Federico and Boen, Joseph and Liu, Sheng and Lu, Pan and Huang, Zhi and Guestrin, Carlos and Zou, James},
  title = {Optimizing Generative {AI} by Backpropagating Language Model Feedback},
  journal = {Nature},
  volume = {639},
  number = {8055},
  pages = {609--616},
  year = {2025},
  doi = {10.1038/s41586-025-08661-4},
  url = {https://www.nature.com/articles/s41586-025-08661-4}
}

@article{skillsbench2026,
  author        = {Xiangyi Li and Yimin Liu and Wenbo Chen and Bingran You and Zonglin Di and Yifeng He and Shenghan Zheng and Kyoung Whan Choe and Jiankai Sun and Shuyi Wang and others},
  title         = {{SkillsBench: Benchmarking How Well Agent Skills Work Across Diverse Tasks}},
  journal       = {arXiv preprint arXiv:2602.12670},
  note          = {Version 4, June 2026},
  year          = {2026},
  eprint        = {2602.12670v4},
  archivePrefix = {arXiv},
  primaryClass  = {cs.AI},
  url           = {https://arxiv.org/abs/2602.12670v4}
}

@article{legoues2012genprog,
  author = {Le Goues, Claire and Nguyen, ThanhVu and Forrest, Stephanie and Weimer, Westley},
  title = {{GenProg}: A Generic Method for Automatic Software Repair},
  journal = {IEEE Transactions on Software Engineering},
  year = {2012},
  volume = {38},
  number = {1},
  pages = {54--72},
  doi = {10.1109/TSE.2011.104},
  url = {https://clairelegoues.com/assets/papers/genprog12tse.pdf}
}

@inproceedings{smith2015overfitting,
  author = {Smith, Edward K. and Barr, Earl T. and Le Goues, Claire and Brun, Yuriy},
  title = {Is the Cure Worse Than the Disease? Overfitting in Automated Program Repair},
  booktitle = {Proceedings of the 2015 10th Joint Meeting on Foundations of Software Engineering},
  year = {2015},
  pages = {532--543},
  publisher = {Association for Computing Machinery},
  doi = {10.1145/2786805.2786825},
  url = {https://people.cs.umass.edu/~brun/pubs/pubs/Smith15fse.pdf}
}

@inproceedings{qi2015test,
  author = {Qi, Zichao and Long, Fan and Achour, Sara and Rinard, Martin},
  title = {An Analysis of Patch Plausibility and Correctness for Generate-and-Validate Patch Generation Systems},
  booktitle = {Proceedings of the 2015 International Symposium on Software Testing and Analysis},
  year = {2015},
  pages = {24--36},
  publisher = {Association for Computing Machinery},
  doi = {10.1145/2771783.2771791},
  url = {https://people.csail.mit.edu/rinard/paper/issta15.pdf}
}

@inproceedings{jones2002spectrum,
  author = {Jones, James A. and Harrold, Mary Jean and Stasko, John},
  title = {Visualization of Test Information to Assist Fault Localization},
  booktitle = {Proceedings of the 24th International Conference on Software Engineering},
  year = {2002},
  pages = {467--477},
  publisher = {Association for Computing Machinery},
  doi = {10.1145/581339.581397}
}

@inproceedings{bouzenia2024repairagent,
  author    = {Islem Bouzenia and Premkumar Devanbu and Michael Pradel},
  title     = {{RepairAgent: An Autonomous, LLM-Based Agent for Program Repair}},
  booktitle = {Proceedings of the 47th IEEE/ACM International Conference on Software Engineering},
  series    = {ICSE '25},
  year      = {2025},
  pages     = {2188--2200},
  publisher = {IEEE},
  doi       = {10.1109/ICSE55347.2025.00157}
}

@article{xia2024agentless,
  author  = {Chunqiu Steven Xia and Yinlin Deng and Soren Dunn and Lingming Zhang},
  title   = {Demystifying {LLM}-Based Software Engineering Agents},
  journal = {Proceedings of the ACM on Software Engineering},
  volume  = {2},
  number  = {FSE},
  year    = {2025},
  pages   = {801--824},
  doi     = {10.1145/3715754}
}

@inproceedings{orso2003leveraging,
  author    = {Alessandro Orso and Taweesup Apiwattanapong and Mary Jean Harrold},
  title     = {Leveraging Field Data for Impact Analysis and Regression Testing},
  booktitle = {Proceedings of the 9th European Software Engineering Conference Held Jointly with the 11th ACM SIGSOFT International Symposium on Foundations of Software Engineering},
  series    = {ESEC/FSE '03},
  year      = {2003},
  pages     = {128--137},
  publisher = {Association for Computing Machinery},
  doi       = {10.1145/940071.940089}
}

@inproceedings{fast2010designing,
  author    = {Ethan Fast and Claire Le Goues and Stephanie Forrest and Westley Weimer},
  title     = {Designing Better Fitness Functions for Automated Program Repair},
  booktitle = {Proceedings of the 12th Annual Conference on Genetic and Evolutionary Computation},
  series    = {GECCO '10},
  year      = {2010},
  pages     = {965--972},
  publisher = {Association for Computing Machinery},
  doi       = {10.1145/1830483.1830654}
}

@article{lou2024rtsapr,
  author  = {Yiling Lou and Jun Yang and Samuel Benton and Dan Hao and Lin Tan and Zhenpeng Chen and Lu Zhang and Lingming Zhang},
  title   = {When Automated Program Repair Meets Regression Testing---An Extensive Study on Two Million Patches},
  journal = {ACM Transactions on Software Engineering and Methodology},
  volume  = {33},
  number  = {7},
  year    = {2024},
  pages   = {1--23},
  doi     = {10.1145/3672450}
}

@inproceedings{jimenez2024swebench,
  author    = {Carlos E. Jimenez and John Yang and Alexander Wettig and Shunyu Yao and Kexin Pei and Ofir Press and Karthik R. Narasimhan},
  title     = {{SWE}-bench: Can Language Models Resolve Real-world {GitHub} Issues?},
  booktitle = {The Twelfth International Conference on Learning Representations},
  series    = {ICLR '24},
  year      = {2024},
  url       = {https://openreview.net/forum?id=VTF8yNQM66}
}

@inproceedings{wang2024agentworkflow,
  author = {Wang, Zora Zhiruo and Mao, Jiayuan and Fried, Daniel and Neubig, Graham},
  title = {Agent Workflow Memory},
  booktitle = {Proceedings of the 42nd International Conference on Machine Learning},
  series = {Proceedings of Machine Learning Research},
  volume = {267},
  pages = {63897--63911},
  year = {2025},
  publisher = {PMLR},
  url = {https://proceedings.mlr.press/v267/wang25bx.html}
}

@book{aho2006compilers,
  author = {Aho, Alfred V. and Lam, Monica S. and Sethi, Ravi and Ullman, Jeffrey D.},
  title = {Compilers: Principles, Techniques, and Tools},
  edition = {2},
  year = {2006},
  publisher = {Addison-Wesley},
  isbn = {9780321486813},
  url = {https://www.pearson.com/en-us/subject-catalog/p/compilers-principles-techniques-and-tools/P200000003472/9780321486813}
}

@article{ferrante1987pdg,
  author = {Ferrante, Jeanne and Ottenstein, Karl J. and Warren, Joe D.},
  title = {The Program Dependence Graph and Its Use in Optimization},
  journal = {ACM Transactions on Programming Languages and Systems},
  volume = {9},
  number = {3},
  pages = {319--349},
  year = {1987},
  doi = {10.1145/24039.24041},
  url = {https://research.ibm.com/publications/the-program-dependence-graph-and-its-use-in-optimization}
}

@inproceedings{mathew2021docvqa,
  author = {Mathew, Minesh and Karatzas, Dimosthenis and Jawahar, C. V.},
  title = {{DocVQA}: A Dataset for {VQA} on Document Images},
  booktitle = {Proceedings of the IEEE/CVF Winter Conference on Applications of Computer Vision},
  pages = {2200--2209},
  month = jan,
  year = {2021},
  doi = {10.1109/WACV48630.2021.00225},
  url = {https://openaccess.thecvf.com/content/WACV2021/html/Mathew_DocVQA_A_Dataset_for_VQA_on_Document_Images_WACV_2021_paper.html}
}

@inproceedings{biten2019stvqa,
  author = {Biten, Ali Furkan and Tito, Ruben and Mafla, Andres and Gomez, Lluis and Rusinol, Marcal and Valveny, Ernest and Jawahar, C. V. and Karatzas, Dimosthenis},
  title = {Scene Text Visual Question Answering},
  booktitle = {Proceedings of the IEEE/CVF International Conference on Computer Vision},
  pages = {4291--4301},
  month = oct,
  year = {2019},
  doi = {10.1109/ICCV.2019.00439},
  url = {https://openaccess.thecvf.com/content_ICCV_2019/html/Biten_Scene_Text_Visual_Question_Answering_ICCV_2019_paper.html}
}

@article{he2026livemathematicianbench,
  author = {He, Linyang and Yu, Qiyao and Dong, Hanze and Liao, Baohao and Xu, Xinxing and Goldblum, Micah and Bian, Jiang and Mesgarani, Nima},
  title = {{LiveMathematicianBench}: A Live Benchmark for Mathematician-Level Reasoning with Proof Sketches},
  journal = {arXiv preprint arXiv:2604.01754v1},
  year = {2026},
  doi = {10.48550/arXiv.2604.01754},
  eprint = {2604.01754v1},
  archivePrefix = {arXiv},
  primaryClass = {cs.CL},
  url = {https://arxiv.org/abs/2604.01754v1}
}

@article{rothermel1997safe,
  author = {Rothermel, Gregg and Harrold, Mary Jean},
  title = {A Safe, Efficient Regression Test Selection Technique},
  journal = {ACM Transactions on Software Engineering and Methodology},
  volume = {6},
  number = {2},
  pages = {173--210},
  year = {1997},
  doi = {10.1145/248233.248262}
}

@article{lin2026wml,
  title = {{Workflow-Localized Mechanism Learning: Attribution-Guided Repair and Knowledge Reuse for Structured Agent Skills}},
  author = {Lin, Zibin and Zhang, Shengli and Wang, Taotao and Xia, Yihan and Ma, Deen and Liao, Guofu},
  journal = {arXiv preprint arXiv:2607.20999},
  year = {2026},
  url = {https://arxiv.org/abs/2607.20999}
}

@article{lin2026skillboost,
  title = {{Rethinking Self-Evolution: A Constrained Exploration-Exploitation Process for Mitigating Skill Overfitting}},
  author = {Lin, Hongqiang and Liu, Chao and Bai, Xiaofan and Jin, Xuan and Li, Yuhong and Zheng, Nenggan and Cao, Xipeng},
  journal = {arXiv preprint arXiv:2607.26643},
  year = {2026},
  url = {https://arxiv.org/abs/2607.26643}
}

@article{sun2026graphskillevo,
  title = {{GraphSkillEvo: Evolutionary Optimization of Graph-Structured Agent Skills}},
  author = {Sun, Rui and Zheng, Zhi and Wang, Zhenkun and Lu, Zhichao},
  journal = {arXiv preprint arXiv:2609.21749},
  year = {2026},
  url = {https://arxiv.org/abs/2609.21749}
}

@article{tan2026skillzip,
  title = {{SkillZip: Contract-Preserving Graph Compression for Scalable Agent Skill Libraries}},
  author = {Tan, Xingyu and Wang, Xiaoyang and Liu, Qing and Xu, Xiwei and Yuan, Xin and Zhu, Liming and Zhang, Wenjie},
  journal = {arXiv preprint arXiv:2608.05604},
  year = {2026},
  url = {https://arxiv.org/abs/2608.05604}
}

@article{wang2024voyager,
  title = {{Voyager: An Open-Ended Embodied Agent with Large Language Models}},
  author = {Wang, Guanzhi and Xie, Yuqi and Jiang, Yunfan and Mandlekar, Ajay and Xiao, Chaowei and Zhu, Yuke and Fan, Linxi and Anandkumar, Anima},
  journal = {Transactions on Machine Learning Research},
  year = {2024},
  url = {https://arxiv.org/abs/2305.16291}
}

@inproceedings{zhou2023ape,
  title = {Large Language Models Are Human-Level Prompt Engineers},
  author = {Zhou, Yongchao and Muresanu, Andrei Ioan and Han, Ziwen and Paster, Keiran and Pitis, Silviu and Chan, Harris and Ba, Jimmy},
  booktitle = {International Conference on Learning Representations},
  year = {2023},
  url = {https://arxiv.org/abs/2211.01910}
}

@inproceedings{pryzant2023protegi,
  title = {Automatic Prompt Optimization with ``Gradient Descent'' and Beam Search},
  author = {Pryzant, Reid and Iter, Dan and Li, Jerry and Lee, Yin Tat and Zhu, Chenguang and Zeng, Michael},
  booktitle = {Proceedings of the 2023 Conference on Empirical Methods in Natural Language Processing},
  year = {2023},
  url = {https://arxiv.org/abs/2305.03495}
}

@inproceedings{khattab2024dspy,
  title = {{DSPy: Compiling Declarative Language Model Calls into Self-Improving Pipelines}},
  author = {Khattab, Omar and Singhvi, Arnav and Maheshwari, Paridhi and Zhang, Zhiyuan and Santhanam, Keshav and Vardhamanan, Sri and Haq, Saiful and Sharma, Ashutosh and Joshi, Thomas T. and Moazam, Hanna and Miller, Heather and Zaharia, Matei and Potts, Christopher},
  booktitle = {International Conference on Learning Representations},
  year = {2024},
  url = {https://arxiv.org/abs/2310.03714}
}

@inproceedings{agrawal2026gepa,
  title = {{GEPA: Reflective Prompt Evolution Can Outperform Reinforcement Learning}},
  author = {Agrawal, Lakshya A. and Tan, Shangyin and Soylu, Dilara and Ziems, Noah and Khare, Rishi and Opsahl-Ong, Krista and Singhvi, Arnav and Shandilya, Herumb and Ryan, Michael J. and Jiang, Meng and Potts, Christopher and Sen, Koushik and Dimakis, Alexandros G. and Stoica, Ion and Klein, Dan and Zaharia, Matei and Khattab, Omar},
  booktitle = {International Conference on Learning Representations},
  year = {2026},
  url = {https://arxiv.org/abs/2507.19457}
}

@inproceedings{cheng2024trace,
  title = {Trace is the Next AutoDiff: Generative Optimization with Rich Feedback, Execution Traces, and {LLMs}},
  author = {Cheng, Ching-An and Nie, Allen and Swaminathan, Adith},
  booktitle = {Advances in Neural Information Processing Systems},
  year = {2024},
  url = {https://arxiv.org/abs/2406.16218}
}

@inproceedings{schnabel2024sammo,
  title = {Symbolic Prompt Program Search: A Structure-Aware Approach to Efficient Compile-Time Prompt Optimization},
  author = {Schnabel, Tobias and Neville, Jennifer},
  booktitle = {Findings of the Association for Computational Linguistics: EMNLP 2024},
  pages = {670--686},
  year = {2024},
  url = {https://arxiv.org/abs/2404.02319}
}

@article{sharma2026mpo,
  title = {Modular Prompt Optimization: Optimizing Structured Prompts with Section-Local Textual Gradients},
  author = {Sharma, Prith and Henley, Austin Z.},
  journal = {arXiv preprint arXiv:2601.04055},
  year = {2026},
  url = {https://arxiv.org/abs/2601.04055}
}

@inproceedings{ma2026sepo,
  title = {{SEPO: Evidence-Grounded Prompt Optimization via Structural Editing}},
  author = {Ma, Xiaoyu and Liu, Haoyue and Li, Yiwen and Zhu, Jionghao and Wang, Zhichao and Chen, Ye and Tang, Xiaoying},
  booktitle = {Findings of the Association for Computational Linguistics: EMNLP 2026},
  year = {2026},
  url = {https://arxiv.org/abs/2608.28067}
}

@inproceedings{zhang2026ace,
  title = {Agentic Context Engineering: Evolving Contexts for Self-Improving Language Models},
  author = {Zhang, Qizheng and Hu, Changran and Upasani, Shubhangi and Ma, Boyuan and Hong, Fenglu and Kamanuru, Vamsidhar and Rainton, Jay and Wu, Chen and Ji, Mengmeng and Li, Hanchen and Thakker, Urmish and Zou, James and Olukotun, Kunle},
  booktitle = {International Conference on Learning Representations},
  year = {2026},
  url = {https://arxiv.org/abs/2510.04618}
}

@article{liang2025promptsprograms,
  title = {Prompts Are Programs Too! Understanding How Developers Build Software Containing Prompts},
  author = {Liang, Jenny T. and Lin, Melissa and Rao, Nikitha and Myers, Brad A.},
  journal = {Proceedings of the ACM on Software Engineering},
  volume = {2},
  number = {FSE},
  year = {2025},
  url = {https://arxiv.org/abs/2409.12447}
}

@article{sharma2025promptpex,
  title = {{PromptPex: Automatic Test Generation for Language Model Prompts}},
  author = {Sharma, Reshabh K. and De Halleux, Jonathan and Barke, Shraddha and Grossman, Dan and Zorn, Benjamin},
  journal = {arXiv preprint arXiv:2503.05070},
  year = {2025},
  url = {https://arxiv.org/abs/2503.05070}
}

@article{sharma2026contextcov,
  title = {{ContextCov: Deriving and Enforcing Executable Constraints from Agent Instruction Files}},
  author = {Sharma, Reshabh K.},
  journal = {arXiv preprint arXiv:2603.00822},
  year = {2026},
  url = {https://arxiv.org/abs/2603.00822}
}

@article{ouyang2025nondeterminism,
  title = {An Empirical Study of the Non-Determinism of {ChatGPT} in Code Generation},
  author = {Ouyang, Shuyin and Zhang, Jie M. and Harman, Mark and Wang, Meng},
  journal = {ACM Transactions on Software Engineering and Methodology},
  volume = {34},
  number = {2},
  year = {2025},
  doi = {10.1145/3697010},
  url = {https://arxiv.org/abs/2308.02828}
}

@misc{anthropic2025skills,
  author = {{Anthropic}},
  title = {Introducing Agent Skills},
  year = {2025},
  howpublished = {\url{https://claude.com/blog/skills}},
  note = {Published 16 October 2025; updated 18 December 2025. Accessed 18 October 2026}
}

@misc{skillssh2026,
  author = {{Vercel}},
  title = {{skills.sh}: The Agent Skills Directory},
  year = {2026},
  howpublished = {\url{https://skills.sh}},
  note = {All-time installation leaderboard lists 1,489,086 skills. Accessed 24 October 2026}
}

@misc{agentskills2025spec,
  author = {{Agent Skills}},
  title = {Agent Skills: Overview and Specification},
  year = {2025},
  howpublished = {\url{https://agentskills.io}},
  note = {Accessed 18 October 2026}
}

@article{dong2026harmful,
  title = {{Agent Skills Can Be Harmful: An Empirical Study of Skill-Induced Failures in LLM Agents}},
  author = {Dong, Gen and Gao, Yanjie and Li, Liqun and Xu, Tianyin and Hua, Yu and Yang, Fan},
  journal = {arXiv preprint arXiv:2608.11888},
  year = {2026},
  url = {https://arxiv.org/abs/2608.11888}
}

@article{tank2026regression,
  title = {{The Regression Tax: Decomposing Why Skills Help and Hurt LLM Agents}},
  author = {Tank, Darshan and Nama, Baran},
  journal = {arXiv preprint arXiv:2607.22520},
  year = {2026},
  url = {https://arxiv.org/abs/2607.22520}
}

@inproceedings{moonen2001island,
  title = {Generating Robust Parsers using Island Grammars},
  author = {Moonen, Leon},
  booktitle = {Proceedings of the 8th Working Conference on Reverse Engineering},
  pages = {13--22},
  publisher = {IEEE Computer Society},
  year = {2001},
  doi = {10.1109/WCRE.2001.957806}
}

@inproceedings{dagenais2008partial,
  title = {Enabling Static Analysis for Partial {Java} Programs},
  author = {Dagenais, Barth{\'e}l{\'e}my and Hendren, Laurie},
  booktitle = {Proceedings of the 23rd ACM SIGPLAN Conference on Object-Oriented Programming Systems, Languages and Applications},
  pages = {313--328},
  year = {2008},
  doi = {10.1145/1449764.1449790}
}

@article{foster2007lenses,
  title = {Combinators for Bidirectional Tree Transformations: A Linguistic Approach to the View-Update Problem},
  author = {Foster, J. Nathan and Greenwald, Michael B. and Moore, Jonathan T. and Pierce, Benjamin C. and Schmitt, Alan},
  journal = {ACM Transactions on Programming Languages and Systems},
  volume = {29},
  number = {3},
  year = {2007},
  articleno = {17},
  doi = {10.1145/1232420.1232424},
  url = {https://doi.org/10.1145/1232420.1232424}
}

@article{weiser1984slicing,
  title = {Program Slicing},
  author = {Weiser, Mark},
  journal = {IEEE Transactions on Software Engineering},
  volume = {SE-10},
  number = {4},
  pages = {352--357},
  year = {1984},
  doi = {10.1109/TSE.1984.5010248}
}

@article{horwitz1990slicing,
  title = {Interprocedural Slicing Using Dependence Graphs},
  author = {Horwitz, Susan and Reps, Thomas and Binkley, David},
  journal = {ACM Transactions on Programming Languages and Systems},
  volume = {12},
  number = {1},
  pages = {26--60},
  year = {1990},
  doi = {10.1145/77606.77608}
}

@article{binkley1997regression,
  author = {Binkley, David},
  title = {Semantics Guided Regression Test Cost Reduction},
  journal = {IEEE Transactions on Software Engineering},
  volume = {23},
  number = {8},
  pages = {498--516},
  year = {1997},
  doi = {10.1109/32.624306},
  url = {https://www.cs.loyola.edu/~binkley/papers/tse97.ps.gz}
}

@article{cawley2010overfitting,
  author = {Cawley, Gavin C. and Talbot, Nicola L. C.},
  title = {On Over-fitting in Model Selection and Subsequent Selection Bias in Performance Evaluation},
  journal = {Journal of Machine Learning Research},
  volume = {11},
  number = {70},
  pages = {2079--2107},
  year = {2010},
  url = {https://www.jmlr.org/papers/v11/cawley10a.html}
}

@article{livshits2015soundiness,
  author = {Livshits, Benjamin and Sridharan, Manu and Smaragdakis, Yannis and Lhot{\'a}k, Ond{\v{r}}ej and Amaral, J. Nelson and Chang, Bor-Yuh Evan and Guyer, Samuel Z. and Khedker, Uday P. and M{\o}ller, Anders and Vardoulakis, Dimitrios},
  title = {In Defense of Soundiness: A Manifesto},
  journal = {Communications of the ACM},
  volume = {58},
  number = {2},
  pages = {44--46},
  year = {2015},
  doi = {10.1145/2644805}
}

@inproceedings{goguen1982security,
  author = {Goguen, Joseph A. and Meseguer, Jos{\'e}},
  title = {Security Policies and Security Models},
  booktitle = {Proceedings of the 1982 IEEE Symposium on Security and Privacy},
  pages = {11--20},
  year = {1982},
  doi = {10.1109/SP.1982.10014}
}

@article{vandeursen1993origin,
  author  = {van Deursen, Arie and Klint, Paul and Tip, Frank},
  title   = {Origin Tracking},
  journal = {Journal of Symbolic Computation},
  volume  = {15},
  number  = {5--6},
  pages   = {523--545},
  year    = {1993},
}

@article{beurerkellner2023lmql,
  author  = {Luca Beurer-Kellner and Marc Fischer and Martin Vechev},
  title   = {Prompting Is Programming: A Query Language for Large Language Models},
  journal = {Proceedings of the ACM on Programming Languages},
  volume  = {7},
  number  = {PLDI},
  year    = {2023},
  doi     = {10.1145/3591300}
}

\appendix
\section{Implementation Details} \label{app:impl}

\chong{This appendix gives the formal definitions and proofs that the main text summarizes, implementation details of \method, the record of the running example, and further analyses of repair outcomes.}

\subsection{Formal Definitions} \label{app:formal}

This subsection restates the definitions of Section~\ref{sec:representation} with all components named.
A span $[\alpha,\beta)$ with $0\le\alpha\le\beta\le|S|$ denotes the substring $S[\alpha,\beta)$ of a skill document $S$.
We write $[\alpha,\beta)\subseteq[\gamma,\eta)$ if $\gamma\le\alpha$ and $\beta\le\eta$, and the two spans are disjoint if $\beta\le\gamma$ or $\eta\le\alpha$.

\parhead{\skir} \chong{A \skir is a tuple $I=(N,P,\mathit{own},\mathit{sec},\mathit{st},\mathit{attr},\mathit{Def},\mathit{Use},\mathit{gov},\mathit{mg})$, where $\mathit{gov}$ and $\mathit{mg}(b)=m_b$ are the maps that the compiler establishes below and $N$ is a finite set of instructions}, each with text $\mathit{txt}(n)$, which retains in full the content not represented in $P$; $P=P_{\mathrm{op}}\uplus P_{\mathrm{br}}\uplus P_{\mathrm{mg}}$ is a finite set of program elements, namely operations, branch conditions, and merge points; $\mathit{own}:P\to N$ maps each element to the instruction from which it was extracted; $\mathit{sec}(n)$ is the path of sections that contain $n$; $\mathit{st}:N\to\{\textsf{full},\textsf{partial}\}$ marks instructions whose requirements are only partly represented by elements of $P$; $\mathit{attr}(n)$ is a finite map from attribute names, such as modality and trigger, to the values that the compiler extracts from $\mathit{txt}(n)$; and $\mathit{Def},\mathit{Use}:P\to 2^{Z}$ give the data items, drawn from a finite set $Z$ of recognized data references, that an element defines and uses, with $\mathit{Def}(p)=\varnothing$ for $p\notin P_{\mathrm{op}}$.
\chong{For each branch condition $b\in P_{\mathrm{br}}$ that a guard recognizer identifies, the compiler establishes, from instruction boundaries and document order, its merge point $m_b\in P_{\mathrm{mg}}$ and the operations $\mathit{gov}(b)\subseteq P_{\mathrm{op}}$ that it governs.}

\parhead{Source map} Let \chong{$\rho(\,\cdot\,;\theta_n)$} denote the parser's projection from an instruction's source span to its text, which removes Markdown syntax such as list markers and line breaks according to the parsing rule \chong{$\theta_n$} that produced the instruction.
Let $P_{\mathrm{src}}\subseteq P$ \chong{contain the program elements that the compiler anchors to a span of text.}
\chong{A source map from $I$ to $S$ is a function $M$ from $N\cup P_{\mathrm{src}}$ to the spans of $S$ such that} (i)~$\mathit{txt}(n)=\rho(S[M(n)];\theta_n)$ for every $n\in N$; (ii)~$M(n)$ and $M(n')$ are disjoint for distinct $n,n'\in N$; and \chong{(iii)~$M(p)\subseteq M(\mathit{own}(p))$ for every $p\in P_{\mathrm{src}}$.}

\parhead{\chong{\skast construction}} \chong{\skast is built from the recognized elements by four rules.} (1)~\juan{A block is an ordered container whose statements are instructions, recognized branch conditions, or nested blocks.
Instructions and conditions are translated by rules (2)--(4), and nested blocks are translated recursively by this rule.
The block becomes a \textsf{scope} whose body sequences the resulting terms in source order.} (2)~The operations of an instruction form a sequence in the order of their source anchors. (3)~\juan{For each branch condition $b$, recognized source syntax determines arm membership. Operations governed by a positive guard such as \cue{if} $b$ or \cue{when} $b$ enter the true arm, whereas those governed by \cue{unless} $b$ or a recognized \cue{else} region of the same branch enter the false arm.
A branch condition $b$ becomes $\textsf{if}\ b\ \textsf{then}\ \pi_b^{\mathsf{t}}\ \textsf{else}\ \pi_b^{\mathsf{f}}$, where $\pi_b^{\mathsf{t}}$ and $\pi_b^{\mathsf{f}}$ sequence the operations in $\mathit{gov}(b)$ assigned to the true and false arms, respectively.
If no operations are recognized for an arm, its term is $\epsilon$.
These operations are placed within their corresponding arms and removed from the surrounding sequence, and the \textsf{if} takes the position of $b$ and is closed by its merge point.} (4)~An instruction without recognized elements contributes $\epsilon$.

\subsection{Compilation} \label{app:compile}

\chong{This subsection describes the identifiers, recognition rules, data references, source maps, and sentence splitting of the compiler and closes with a table of its compilation rules.
\parhead{Identifiers} Each instruction receives a sequential parsing identifier and a canonical identifier.}
The canonical identifier is formed by prefixing \texttt{S} to the first 16 hexadecimal characters of a SHA-256 digest over the normalized section path, inferred instruction type, normalized instruction text, and duplicate-occurrence index.
\chong{Absolute source offsets do not enter this identifier, so an earlier edit that merely shifts an instruction's span leaves its canonical identity unchanged,} whereas changing its normalized text, section, type, or duplicate index may change it.
\chong{Labels that the initial skill gives to instructions, such as \texttt{D010}, are retained separately as external provenance.}

\parhead{Action and condition recognition} \juan{Action recognition uses a fixed, word-bounded lexicon over the complete instruction text.
\chong{The lexicon includes source verbs such as \cue{read} and \cue{load}, selection verbs such as \cue{find} and \cue{choose},} checking and transformation verbs such as \cue{validate} and \cue{normalize}, and output or mutation verbs \chong{such as \cue{return} and \cue{patch}.
Recognized occurrences produce operations in their order in the text.}
Instructions beginning with \cue{if}, \cue{when}, \cue{unless}, \cue{in case}, or \cue{for} are classified as conditional.
The first three forms additionally expose a trigger when the condition ends at a comma, colon, or \cue{then}. }
\juan{The compiler also recognizes postposed \cue{provided that} and \cue{only if} clauses, whose conditions govern the preceding action.}
Loop-related words such as \cue{repeat}, \cue{until}, and \cue{while} \juan{also affect instruction classification.}
The authored loop-related requirements remain available in the retained instruction text.
Table~\ref{tab:ir-elements} lists the element types that these rules produce.

\begin{table}[!htbp]
\centering
\caption{How the compiler represents supported skill content. Text that
cannot be fully parsed is preserved and marked as partially parsed.}
\label{tab:ir-elements}
\papertablestyle
\begin{tabular}{@{}ll@{}}
\toprule
Skill content & \skir element \\
\midrule
An instruction & \chong{instruction ($N$)} \\
An explicit operation & \chong{operation ($P_{\mathrm{op}}$)} \\
An explicit condition & \chong{branch condition ($P_{\mathrm{br}}$)} \\
A branch merge & \chong{merge point ($P_{\mathrm{mg}}$)} \\
A recognized data reference &
\chong{data item / def--use relation} \\
Unresolved content &
\chong{instruction text with status \textsf{partial}} \\
\bottomrule
\end{tabular}
\end{table}

\parhead{Conditions and data references} A \cue{When to use:} \chong{clause conditions the immediately following prose instruction in the same section if it contains a recognized action verb.}
\chong{The clause instruction owns the branch condition and its merge point, and the condition is source-mapped to the clause.}
A control edge connects the branch to the governed instruction without changing that instruction's stored trigger or predicate.
\chong{During compilation, each use of a data item is linked to the most recent preceding definition of the same normalized item in document order.}
An operation's uses are processed before its outputs update that producer.
The IR-to-AST pass assigns a separate variable to each producer definition and uses these relations to select call arguments.
The final def--use edges in $A_{\mathrm{ddg}}$ are then generated by fixed-point reaching-definitions analysis over the CFG.
\chong{This analysis joins definitions of the same variable across predecessors, where the pass that builds \skast from \skir gives each definition of a data item its own variable.}

\parhead{Source maps and well-formedness} In addition to offsets, each mapping records the source file, a hash of the source text, and line and column positions.
\chong{Materialization replaces only the mapped span and leaves all surrounding text, including sibling sentences and Markdown formatting, unchanged byte for byte.}
For the repair of Section~\ref{sec:overview-walkthrough}, \chong{recompilation recovers the replacement text, so target recovery holds, and maps the rewritten instruction to its source span.}
\chong{The well-formedness checks verify that every address of the address contract and every data reference resolve, and that the source map satisfies the definition of a source map for the compiled document.}
\chong{Source-map integrity checks validate the source text and hashes, instruction-span bounds and stored offsets, and the links between the canonical identifier, parsing identifier, and imported label of each instruction.}

\parhead{Sentence splitting and status} Sentence splitting determines the instruction units passed to the recognizers.
For ordinary single-line prose, the parser splits a list item or paragraph when conservative punctuation rules, excluding recognized abbreviations, identify at least two nonempty sentence spans.
Otherwise, it retains the entire item or paragraph, as it does for multiline prose, code blocks, and tables.
In either case, the resulting instructions retain source spans that anchor them in the document.
The compiler then applies its recognizers to each instruction and assigns its status from the semantic evidence found.
If no explicit semantic evidence is recognized, the instruction is marked \textsf{partial}, while its authored text remains available through its source span in the partial program view.

\parhead{Source-map granularity} Instruction mappings follow the parser's granularity.
Sentence-split instructions receive exact, disjoint sentence spans, for which the projection is the identity, and unsplit instructions retain the source-line intervals consumed by the parser, including any list markers and terminal line separators.
Program-element mappings may identify exact textual spans or \chong{\chong{coarser source anchors, which need not denote independently authored text.}}
The metadata $\theta_n$ describes the parser's extraction rule and is not an additional field stored in the source map.

Table~\ref{tab:compile-rules} summarizes the compilation rules of Section~\ref{sec:representation}.

\begin{table}[!t]
\centering
\caption{\chong{Compilation rules of \skir. Each row gives a textual pattern, what the compiler produces from it, and its effect on $\mathit{Def}$ and $\mathit{Use}$.}}
\label{tab:compile-rules}
\papertablestyle
\small
\setlength{\tabcolsep}{4pt}
\begin{tabularx}{\linewidth}{@{}>{\raggedright\arraybackslash}X>{\raggedright\arraybackslash}p{0.2\linewidth}>{\raggedright\arraybackslash}p{0.2\linewidth}>{\raggedright\arraybackslash}X@{}}
\toprule
Pattern in instruction text & Element & $\mathit{Def}$ / $\mathit{Use}$ & Example \\
\midrule
Word-bounded action from the fixed lexicon &
operation &
Source actions define; checking or transformation actions use and define; other actions use & \chong{\cue{return only the requested value span}} \\

\juan{Line-initial or postposed conditional, or \cue{When to use:}} &
branch condition, merge point &
Both have empty Def/Use sets &
\texttt{When to use: For tables} \\

Loop-related control term &
instruction classification &
No independent loop Def/Use rule &
\texttt{repeat validation} \\

Inline Markdown-code span &
data item &
Role determined by the referencing operation &
\texttt{Validate `report'} \\

Other authored content &
retained instruction text &
No additional extracted Def/Use sets &
The units exception of the value-span instruction \\
\bottomrule
\end{tabularx}
\end{table}

\subsection{Analysis Example} \label{app:analysis-example}

Figure~\ref{fig:analysis-example} illustrates the analyses on a schematic
procedure in which all three relations matter. Each instruction contains a
single element, which we identify with the instruction. We write
$x\leftarrow a(\ldots)$ for an operation $a$ with $\mathit{Def}(a)=\{x\}$,
and, to show all three relations, \chong{give the condition $n_2$ two branches and its merge point $n_5$ an instruction of its own}:
\begin{equation}
\begin{aligned}
n_1 &: x\leftarrow\operatorname{read}(), &
n_2 &: \operatorname{available}\,?, &
n_3 &: y\leftarrow x, \\
n_4 &: y\leftarrow\texttt{N/A}, &
n_5 &: \text{merge}, &
n_6 &: \operatorname{return}(y).
\end{aligned}
\label{eq:analysis-example-nodes}
\end{equation}
In \skast, $n_3$ and $n_4$ are the branches of an \textsf{if} on
$n_2$, followed by $n_6$. Figure~\ref{fig:analysis-example} shows the resulting edges. The CDG connects $n_2$ to $n_3$ and $n_4$, whereas $n_5$
and $n_6$ postdominate $n_2$ and are not control-dependent on it. The DDG
connects the return $n_6$ to the two assignments that may supply $y$. \chong{The condition $n_2$ that chooses between them is two hops away in the dependence graphs and so lies outside $\mathit{Rel}(n_6)$.} For a patch to the fallback $n_4$,
the dependence neighbors are $n_2$ and $n_6$ and the CFG neighbors are
$n_2$ and $n_5$, so $\mathit{Rel}(n_4)=\{n_2,n_4,n_5,n_6\}$, which excludes
the alternative assignment $n_3$.

\subsection{Components} \label{app:components}

Table~\ref{tab:components} lists the components of \method and the procedure, if any, through which the execution agent or the repair model performs them.
\chong{The repair model explains failures, selects affected instructions, names the source failures, control tests, and boundary tests of a patch, and proposes the target, repair operator, and new content of a typed mutation.}
The host validates these choices, expands the selected instructions through one-hop CFG/CDG/DDG relations, and retrieves repair tests through runtime instruction coverage.
Type checking, source-map materialization, recompilation, test classification, and validation-based acceptance follow fixed rules.

\begin{table}[!t]
\centering
\caption{Components of \method and the procedure, if any, through which
the execution agent or the repair model performs them. Steps marked -- are performed by the deterministic host.}
\label{tab:components}
\papertablestyle
\small
\setlength{\tabcolsep}{4pt}
\begin{tabularx}{\linewidth}{@{}>{\raggedright\arraybackslash\leavevmode}p{0.27\linewidth}c>{\raggedright\arraybackslash\leavevmode}X@{}}
\toprule
Step & By & Input $\rightarrow$ output \\
\midrule
Compilation, \skast, CFG, CDG, DDG & -- &
skill $\rightarrow$ \skir, source map, graphs \\

Execution of a test & \Execute & skill, task input $\rightarrow$ answer, \chong{reported addresses} \\

Fault localization & --, \Diagnose &
traces, \skir $\rightarrow$ suspicious instructions with static context; \Diagnose groups failures \\

Repair hypotheses & \Diagnose &
localization context, traces $\rightarrow$ hypotheses anchored to repair targets \\

Patch content & \Propose &
hypothesis, current content of target $\rightarrow$ typed mutation $(e,o,c)$ \\

Affected instructions and boundary tests & \Diagnose, \Propose &
localization context, repair-test traces $\rightarrow$ affected instructions and boundary tests \\

Type check, materialization, recompilation & -- &
mutation, \skir, source map $\rightarrow$ \chong{patched skill} or rejection \\

Round-trip observation & -- &
mutation, rebuilt \skir $\rightarrow$ target recovery and stability \\

Choice of $Q_\Delta$ & -- &
\chong{footprint $X_\Delta$, $\mathit{Rel}$ $\rightarrow$ one-hop related instructions} \\

Test selection, classification, acceptance & -- &
$Q_\Delta$, source failures, model-selected tests, coverage, outcomes
$\rightarrow$ $L(S')$, groups A--D, decision \\
\bottomrule
\end{tabularx}
\end{table}

\subsection{Execution Traces} \label{app:traces}

\chong{This subsection describes the address contract, the usage report that the execution agent returns, how the host resolves it into coverage, and when recorded executions are reused.
The execution agent receives the authored skill together with a version-specific address contract.}
The contract preserves \chong{public provenance labels, such as \texttt{D010} for instructions and \texttt{E101} for dependencies, when available.}
Otherwise, it assigns deterministic public addresses.
\chong{Canonical \skir identifiers are not shown to the execution agent.}
\chong{Alongside its answer, the execution agent is instructed to return a usage report that lists the instruction and dependency addresses it used, optionally supplemented by condition outcomes, operation states, and artifact events.}
The same execution-address contract and \texttt{graph\_usage} schema are used across repair, validation, and held-out tests, and in the matched w/o \skir ablation.
Validation and test traces are not fed back into repair.
Held-out test evaluation occurs only after selection.
\chong{The host resolves instruction addresses against the current \skir.}
\chong{A dependency address names a dependency of \skir, and the host adds the instructions that own its producer and consumer to the coverage.}
Resolved elements are projected through their ownership to obtain $\mathit{cov}(\tau_i)\subseteq N$.
\chong{References are resolved individually, so an unresolved reference does not discard the resolved references of the same trace.}
Traces are bound to the skill and representation versions that produced them.
\chong{Each patched skill is executed on its selected repair tests $L(S')$.}
\chong{A recorded execution is reused only when the skill, its \skir, the test partition, the execution model, the address contract, and the scoring rule match.}

\subsection{Localization and Repair Hypotheses} \label{app:localization}

\chong{This subsection gives the evidence levels of localization, the retrieval tools and repair memory available to the repair model, and the checks on hypotheses and repair boundaries.
\method retains all failures of the current skill on the repair tests.}
\chong{\Diagnose receives a record for every failure and reviews each record exactly once.}
The repair model determines which failures share a mechanism rather than inheriting a host-defined clustering.
\juan{Graph witnesses provide evidence for localizing these failures by linking recorded execution facts to candidate source instructions through program relations.
A \emph{runtime event} is an execution record containing the run and task identifiers, the operation and its status, the record's provenance, and any reported program-element identifier or observed condition, data, or artifact.
The host represents runtime events and program elements as typed graph nodes.
A \emph{registered edge} is a relation stored in the host's graph registry between registered nodes.
To connect the execution evidence to the program analyses, \emph{cross-representation mappings} link corresponding elements in the execution graph, \skast, CFG, and applicable dependence graphs.
The host checks that the witness path is connected by its recorded edges and that these mappings consistently identify the same recognized action and source instruction.
These checks establish the structural consistency of the witness, while two distinct confidence scores assess the strength of its localization evidence.
\emph{Binding confidence} grades the runtime-to-program association according to the event's provenance, whereas \emph{causal confidence} grades the failure-specific evidence supporting the candidate localization, such as a recorded requirement violation or contrasting outcomes.
The localization levels below use these deterministic, rule-based scores to grade evidence from graph witnesses.}
\juan{Activation contrasts---the difference between activation rates on failing and passing repair tests---help organize the evidence.
\chong{\Diagnose considers all distinct repair targets of the instructions retained by localization, rather than committing to the highest-scoring instruction.}
\chong{Strong evidence comprises requirement violations confirmed by the execution agent or complete graph witnesses with binding and causal confidence at least $0.9$.}
Medium evidence includes direct runtime usage, complete graph witnesses meeting the $0.6$ thresholds but not Strong, and verified missing-capability insertion anchors.
\juan{The evidence levels are ordered Weak, Medium, and Strong.
\method uses Medium as the fixed retention threshold and supplies diagnosis with locations that meet or exceed it and have valid source mappings and mutation authorization.}
\chong{For each affected instruction of a hypothesis $h$, the edit context $\Gamma_t(h)$ contains its exact text, source-linked representation,} incident one-hop CFG/CDG/DDG relations, runtime usage statistics, and repair-test executions whose runtime facts overlap the instruction.}

\chong{The repair model can retrieve further failure records through a retrieval tool, \texttt{read\_\allowbreak train\_\allowbreak evidence}, which returns the input, answer, reference answer, trace, the instructions to which the trace resolves, and the related instructions and graph relations.}
Each repair-model request supports one batched retrieval turn.
\chong{After the model selects affected instructions, the host supplies an index of the earlier patches to these instructions.}
The index records original target IDs together with available canonical IDs \chong{rebuilt from the edited source span, so that earlier repair outcomes can be linked to the corresponding instructions across skill versions.}
\chong{Through a second tool call, the model can inspect earlier edits, fixed and broken tests, preserved and unresolved cases, and repair-boundary observations.}
\chong{\chong{The history does not determine the initial grouping of failures.}}

Hypotheses have fields for the proposed mechanism, supporting failure records, and affected canonical instruction identifiers.
\chong{The repair model also returns a review that accounts for every failure exactly once.}
The host verifies that supporting failures belong to the current failure set and \chong{that affected instructions belong to the supplied set of candidate instructions.
Each hypothesis is validated independently, so an invalid hypothesis does not remove valid ones.}
\chong{The suspicious instructions are derived from repair targets that the traces cover, that have valid source mappings, and that admit at least one repair operator, including insertion anchors.}
Each target retains its supporting evidence and source-linked authored instruction.

There is no fixed hypothesis-count limit.
\chong{Each hypothesis that calls for an edit receives a separate call of \Propose, which returns at most one patch for it.}
A proposal may coordinate edits to multiple interacting instructions.
During generation, \chong{the model also supplies a \texttt{repair\_boundary}, which consists of an activation condition,} \chong{the intended behavior change, behavior to preserve, relevant interactions with other instructions, and repair-test expectations labeled} \texttt{repair}, \texttt{boundary}, or \texttt{preserve}.
The host validates the referenced test identities, includes these tests in local screening, and records their paired before/after outcomes for subsequent repair iterations.

\subsection{Patch Construction} \label{app:patch}

\chong{This subsection describes how the host turns a proposal into a typed mutation and materializes it, the preconditions of each repair operator, and the request sent to \Propose.
The host supplies the selected instructions, their one-hop graph context, the repair tests whose traces cover them, the earlier patches to them, and the legal edit options available at each target.}
Using this information, the model chooses an instruction identifier, a supplied operator, and replacement content satisfying that operator's schema.
The host then matches this choice to a precomputed edit option and \chong{constructs the typed mutation, which fixes its canonical target, the current state of that target, its evidence references, and the source span it may change.}
\chong{It checks that the before-state agrees with the current representation and that the replacement satisfies the operator schema.}
\chong{Finally, it} applies the validated mutation to a copy of the current representation and uses the source map to serialize the updated content within the authorized mapped span.
\chong{Materialization constructs $S'=S_t[0,\alpha)\cdot c'\cdot S_t[\beta,|S_t|)$, where $[\alpha,\beta)$ is that span and $c'$ is the serialized replacement, so the prefix and suffix are unchanged.}
The formal \skir models instructions, program elements, and their $\mathit{Def}$ and $\mathit{Use}$ sets.
Within this representation, typed attributes support targeted edits.
\texttt{MODIFY\_ATTRIBUTE} updates an instruction's \texttt{modality} (\texttt{MUST}, \texttt{SHOULD}, \texttt{MAY}, or \texttt{NEVER}), while \texttt{MODIFY\_CONSTRAINT} updates its modality and optionally the \texttt{trigger} of an existing conditional instruction.
\chong{For these edits, the host derives the updated content from the original text by changing only the corresponding marker or condition.}

\chong{The same operator set also supports instruction insertion through \texttt{ADD\_NODE}, at the end of the anchor's mapped span, and deletion through \texttt{REMOVE\_NODE}, over the target's mapped span.}

\chong{For a relation edit, the host designates a resolved source-mapped instruction as the instruction that states the relation and fixes its canonical identifier in the edit option.}
When resolving this anchor from the relation itself, the host uses the recorded source instruction or a uniquely resolved common instruction of its endpoints.
\chong{The model then supplies the manifestation text, with which the host rewrites the designated instruction.}
On this basis, the host constructs $\mathit{Opt}_t(e)$ from the edit operators whose target, source authority, and typed preconditions hold in the current representation.
For instruction edits, \texttt{MODIFY\_NODE} and \texttt{REMOVE\_NODE} require an existing editable instruction with a mapped source span, while \texttt{ADD\_NODE} requires an authorized insertion boundary at a mapped anchor.
\texttt{MODIFY\_ATTRIBUTE} and \texttt{MODIFY\_CONSTRAINT} additionally require a new modality from $\{\texttt{MUST},\texttt{SHOULD},\texttt{MAY},\texttt{NEVER}\}$.
When the instruction has an existing condition representation, \texttt{MODIFY\_CONDITION} supports a source-representable trigger change, and \texttt{MODIFY\_CONSTRAINT} may include such a change with its modality update.
For relation edits, every option fixes the manifestation instruction and supplies the permitted endpoint, dependency-kind, and data-item domains.
Within these domains, \texttt{ADD\_RELATION} requires a new relation, whereas \texttt{REMOVE\_RELATION}, \texttt{MODIFY\_RELATION}, and \texttt{CHANGE\_DEPENDENCY} require a uniquely identified existing relation with a matching before-state.
The replacement relation, where applicable, must satisfy the supplied typed domains, and its manifestation text must be attached to the fixed instruction.

Table~\ref{tab:operations} lists the repair operators, and Figure~\ref{fig:patch-prompt} shows the request for one call of \Propose.

\begin{table}[!t]
\centering
\caption{\chong{Repair operators. Each row gives the operators, the kind of target they apply to (an instruction, a branch condition, a dependency, or an insertion anchor), the new content they require, and their effect on \skir and on the skill text.}}
\label{tab:operations}
\papertablestyle
\small
\setlength{\tabcolsep}{4pt}
\begin{tabularx}{\linewidth}{
@{}>{\raggedright\arraybackslash}p{0.22\linewidth}
>{\raggedright\arraybackslash}p{0.10\linewidth}
>{\raggedright\arraybackslash}p{0.18\linewidth}
>{\raggedright\arraybackslash}X@{}}
\toprule
Operator & Target & New content &
Effect on \skir and on the text \\
\midrule
\texttt{MODIFY\_\allowbreak NODE} &
\textsf{Instr} &
instruction text &
replaces the instruction's text and recompiles its extracted elements \\
\texttt{MODIFY\_\allowbreak ATTRIBUTE} &
\textsf{Instr} &
modality &
changes the instruction's modality and rewrites its authored
modality expression \\
\texttt{MODIFY\_\allowbreak CONDITION} &
\textsf{Cond} &
condition text &
rewrites an existing instruction trigger and recompiles the
corresponding branch and merge structure \\
\texttt{MODIFY\_\allowbreak CONSTRAINT} &
\textsf{Instr} &
modality, optional trigger &
changes the modality and optionally the trigger
of a conditional instruction \\
\texttt{ADD\_\allowbreak RELATION},
\texttt{REMOVE\_\allowbreak RELATION},
\texttt{MODIFY\_\allowbreak RELATION},
\texttt{CHANGE\_\allowbreak DEPENDENCY} &
\textsf{Dep} &
authored manifestation text; relation specification for
addition or replacement &
adds, removes, or replaces a canonical dependency and rewrites
its designated manifestation instruction \\
\texttt{ADD\_\allowbreak NODE} &
\textsf{Anchor} &
instruction text &
inserts a new instruction after the anchor instruction's
source span \\
\texttt{REMOVE\_\allowbreak NODE} &
\textsf{Instr} &
none &
deletes the instruction and its source span; dependencies
are rebuilt from the resulting text \\
\bottomrule
\end{tabularx}
\end{table}

\begin{figure}[!t]
\centering
\small
\fbox{\begin{minipage}{0.96\linewidth}
\begin{tabularx}{\linewidth}{
@{}>{\sffamily}l>{\raggedright\arraybackslash}X@{}}
Task &
Propose one typed patch for the repair hypothesis, using only the
listed edit options. \\
Hypothesis &
$\langle$failure mechanism$\rangle$,
$\langle$supporting failing repair tests$\rangle$,
$\langle$affected instructions$\rangle$ \\
Instruction &
Relevant instruction text; available local graph facts and
related instruction text; usage statistics on current repair tests;
and legal edit options with their content schemas. \\
Evidence &
$\langle$related failing and passing repair tests of the current skill:
input context, answer, reference, outcome$\rangle$ \\
Memory &
$\langle$node-linked repair memory:
earlier patches and their paired repair-test outcomes$\rangle$ \\
Tools &
\texttt{read\_\allowbreak train\_\allowbreak evidence},
\texttt{read\_repair\_memory}, and
\texttt{inspect\_train\_resources}, as available
(at most one optional batched retrieval turn). \\
Output &
\texttt{node\_id}, \texttt{operation}, and
\texttt{replacement} for each edit, with a required
\texttt{repair\_boundary}: activation condition, \chong{intended change, behavior to preserve, interacting instructions, and a required \texttt{train\_expectations} list, which may be empty.}
Each expectation names a repair test, an expected behavior, and
a role \texttt{repair}, \texttt{boundary}, or \texttt{preserve}. \\
\end{tabularx}
\end{minipage}}
\caption{Structure of the request for one call
$\Propose(h,\Gamma_t(h))$, which carries the edit context of
hypothesis $h$. \chong{Angle brackets mark content that the host fills in from the hypothesis, the traces, and repair memory.}}
\label{fig:patch-prompt}
\Description{A box listing the fields of the patch-generation request:
task, hypothesis, per-instruction context, evidence, repair memory,
retrieval tools, and the output schema.}
\end{figure}

\subsection{Round-Trip Consistency} \label{app:roundtrip}

With explicit offsets, Lemma~\ref{lem:locality} states the following.
Suppose the nonempty intervals in $\mathcal{E}_\Delta$ are pairwise disjoint and no insertion position lies strictly inside one of these intervals.
Let $[\gamma,\eta)$, with $\gamma<\eta$, \chong{be an unedited source fragment, that is, every edit satisfies} $\beta_j\le\gamma$ or $\alpha_j\ge\eta$.
Then $S_t[\gamma,\eta)$ occurs unchanged in $S'$ at $[\gamma+\delta,\eta+\delta)$, where \[ \delta=\sum_{\beta_j\le\gamma} \bigl(|c'_j|-(\beta_j-\alpha_j)\bigr).
\] An insertion at $\gamma$ is included in this shift, whereas one at $\eta$ is not.

\begin{proof}[Proof of Lemma~\ref{lem:locality}] \chong{Each edit interval or insertion position lies entirely before or after the fragment, so by the materialization equation applying that edit copies every character of the fragment unchanged.}
An edit ending at or before $\gamma$ shifts the fragment by $|c'_j|-(\beta_j-\alpha_j)$, and an edit starting at or after $\eta$ does not shift it.
\chong{Materialization applies the edits in descending order of their start offsets and breaks ties by descending end offset, so that a nonempty replacement precedes an insertion at the same position.}
\chong{Under the stated non-overlap conditions, each edit applied later in the descending order ends at or before the start of every edit applied earlier, so each edit is applied at its original source coordinates.}
The resulting shifts therefore compose to give the stated sum.
\end{proof}

\begin{definition}[Round-trip consistency] Let $(I'',M'')=\mathcal{C}(S')$.
The patched skill $S'$ is round-trip consistent if (a)~$(I'',M'')$ passes the well-formedness checks; \chong{(b)~\chong{each mutation passes its operator-specific reconstruction check (\emph{target recovery}), that is, text and attribute edits recover the requested instruction content and fields,} insertions and deletions establish the intended structural change, and relation edits recover the requested relation changes in the recompiled representation; and (c)~each instruction} that is neither rewritten nor removed, including an insertion anchor, retains a counterpart with the same canonical identifier and the same text, modality, trigger, predicate, node type, semantic role, section path, \chong{code-block status, and referenced-assets list (\emph{stability}).}
\chong{Target recovery does not require an edited instruction to keep its canonical identifier.}
\method enforces~(a) before any test runs and records the outcomes of (b) and (c) as round-trip observations.
\end{definition}

Stability does not follow from edit locality alone, \chong{because recompilation also segments the text and assigns each instruction its syntactic class, such as single-line prose, multiline prose, code block, or table, and its section.}
We call $\mathcal{C}$ \emph{segmentation-local for a patch} if every instruction outside $X_\Delta$ is reparsed at its shifted source span with the same boundaries, syntactic class, local source-to-text projection, and section assignment.
Within a fixed syntactic class, an attribute is \emph{context-free} if $\mathcal{C}$ computes it from the instruction's text and section alone.
\chong{Modality, trigger, predicate, instruction type, semantic role, and referenced assets are context-free in this sense.}
Code-block status follows from the syntactic class.
\chong{Canonical identifiers additionally depend on preceding occurrences of the same normalized section, instruction type, and text.}

\begin{lemma}[Stability of context-free attributes] Suppose the source edits satisfy the non-overlap conditions of Edit locality, \chong{every source span $M_t(n)$ for $n\notin X_\Delta$ is an unedited fragment as defined above,} and $\mathcal{C}$ is segmentation-local for the patch.
Then each such instruction has a counterpart in $I''$ with the same text, section, syntactic class, and context-free attributes.
\end{lemma} \begin{proof} Edit locality preserves the characters of $M_t(n)$ at its shifted position.
Segmentation-locality preserves the instruction's boundaries, syntactic class, local source-to-text projection, and section assignment.
Applying that same projection to the preserved characters therefore recovers the same instruction text.
Each context-free attribute is a function of this text and section within the preserved syntactic class, so its value is unchanged.
\end{proof}

The occurrence-dependent canonical identifier remains \chong{subject to the stability check after recompilation, and it does not follow from preservation of the context-free attributes.}

\subsection{Test Selection and Iteration} \label{app:selection}

\chong{This subsection states the properties of Section~\ref{sec:guarantees} as a theorem and proves it together with Proposition~\ref{prop:selection}.}
\chong{To separate what the host decides from what the LLMs return,} Figure~\ref{fig:rules} restates one iteration of Algorithm~\ref{alg:repair} as inference rules.
\chong{Of the premises of the rules, only the results of the three procedures depend on the LLMs.
These are the hypotheses $H$ returned by} \Diagnose in line~3, the patch that \Propose returns for a hypothesis $h\in H$, and the answers of \Execute calls, from which the host computes $R_{\mathrm{target}}(S')$ and $J_{D_{\mathrm{val}}}$.
\textsc{T-Patch} types a mutation against the current \skir.
The target must exist in $I_t$ with the operator's expected kind, the host must offer the operator at that target ($\mathit{Opt}_t(e)$), and the new content must satisfy the operator's schema.
\chong{A coordinated patch $\Delta$ is well typed, written $I_t\vdash\Delta\ \mathsf{ok}$, if $I_t\vdash\Delta_j\ \mathsf{ok}$ holds for each of its mutations $\Delta_j$ and their edits rewrite, delete, or insert after distinct instructions.}
\textsc{Cand} materializes a well-typed patch and requires the recompiled \skir to be well formed, \textsc{Screen} keeps the \chong{patched skills} that fix a source failure, and \textsc{Accept} and \textsc{Keep} choose the next skill from the validation outcomes of the shortlist $\mathit{Top}$ of at most $K$ screened \chong{patched skills}.

\begin{figure}[!t]
\small
\begin{mathpar}
\inferrule*[right=\textsc{T-Patch}]
{e:\kappa\ \text{in}\ I_t \\ o:\kappa\Rightarrow\sigma \\
o\in\mathit{Opt}_t(e) \\ c:\sigma}
{I_t\vdash(e,o,c)\ \mathsf{ok}}
\and
\inferrule*[right=\textsc{Screen}]
{S_t\vdash S'\ \mathsf{cand} \\ R_{\mathrm{target}}(S')>0}
{S_t\vdash S'\ \mathsf{screened}}
\\
\inferrule*[right=\textsc{Cand}]
{h\in H \\ (\Delta,F_\Delta,C_\Delta,B_\Delta)\in\Propose(h,\Gamma_t(h)) \\ I_t\vdash\Delta\ \mathsf{ok} \\
S'=\operatorname{Materialize}(\operatorname{ApplyMutation}(I_t,\Delta),M_t)
\\ \mathcal{C}(S')\ \text{well formed}}
{S_t\vdash S'\ \mathsf{cand}}
\\
\inferrule*[right=\textsc{Accept}]
{S'\in\mathit{Top} \\
J_{D_{\mathrm{val}}}(S')>J_{D_{\mathrm{val}}}(S_t) \\
\forall S''\in\mathit{Top}.\;
J_{D_{\mathrm{val}}}(S'')\le J_{D_{\mathrm{val}}}(S')}
{S_t\Longrightarrow S'}
\and
\inferrule*[right=\textsc{Keep}]
{\forall S'\in\mathit{Top}.\;
J_{D_{\mathrm{val}}}(S')\le J_{D_{\mathrm{val}}}(S_t)}
{S_t\Longrightarrow S_t}
\end{mathpar}
\caption{Rules of one repair iteration. $H$ is the set of
hypotheses returned by $\Diagnose(\Lambda_t)$ that pass the host's
checks, and $\mathit{Top}$ holds at most $K$ skills $S'$ with
$S_t\vdash S'\ \mathsf{screened}$, ranked as described in Section~\ref{sec:selection}\chong{, and \textsc{Accept} takes the first in this order among those with the highest $J_{D_{\mathrm{val}}}$}. $R_{\mathrm{target}}(S')$ and $J_{D_{\mathrm{val}}}$ are recorded outcomes, and
\chong{$J_{D_{\mathrm{val}}}(S_t)$ is the recorded validation pass rate that the iteration receives with $S_t$.}}
\label{fig:rules}
\Description{Five inference rules: T-Patch types a mutation, Cand forms a
well-formed candidate from a well-typed patch, Screen keeps candidates that
fix a source failure, Accept replaces the current skill by the best
shortlisted candidate if it passes more validation tests, and Keep retains
the current skill otherwise.}
\end{figure}

\begin{theorem}[Guarantees of an iteration] \label{thm:iteration} Suppose that $\mathcal{C}(S_t)$ is well formed and $S_t\Longrightarrow S_{t+1}$.
For all results of \Execute, \Diagnose, and \Propose, (i)~$\mathcal{C}(S_{t+1})$ is well formed; (ii)~if \textsc{Accept} applies, $S_{t+1}$ is obtained from $S_t$ by one well-typed patch $\Delta$, and the source span $M_t(n)$ of every instruction $n\notin X_\Delta$ occurs unchanged in $S_{t+1}$, shifted by the edits that precede it; and (iii)~the recorded validation pass rates satisfy $J_{D_{\mathrm{val}}}(S_{t+1})\ge J_{D_{\mathrm{val}}}(S_t)$, strictly if \textsc{Accept} applies.
\end{theorem}

\chong{Part~(ii) rests on Edit locality and on the disjointness of source spans, and the other parts follow from the rules and Algorithm~\ref{alg:repair}.}

\begin{proof}[Proof of Proposition~\ref{prop:selection}] The relation $\mathit{Rel}$ is symmetric.
For $m\neq n$, $m\in\mathit{Rel}(n)$ holds iff some $p$ with $\mathit{own}(p)=n$ and $q$ with $\mathit{own}(q)=m$ satisfy $q\in\mathit{Nbr}(p)$, and $\mathit{Nbr}$ is symmetric by definition, so $n\in\mathit{Rel}(m)$.
For~(i), let $i\notin L(S')$.
Then $i\notin T(X_\Delta\cup Q_\Delta)$, so no instruction in $\mathit{cov}(\tau_i)$ lies in $X_\Delta\cup Q_\Delta$.
Suppose some $u\in X_\Delta$ lies in $\mathit{Ctx}_t(i)$, that is, $u\in\mathit{Rel}(v)$ for some $v\in\mathit{cov}(\tau_i)$.
By symmetry, $v\in\mathit{Rel}(u)\subseteq X_\Delta\cup Q_\Delta$, a contradiction.
Hence $X_\Delta\cap\mathit{Ctx}_t(i)=\varnothing$, and one-hop locality for $\Delta$ gives the claim.
For~(ii), if $i\notin L_0(S')$, then $X_\Delta\cap\mathit{cov}(\tau_i)= \varnothing$, and coverage locality for $\Delta$ gives the claim.
For the negative part, \chong{let $u\in X_\Delta$ be an instruction whose text $\Delta$ rewrites or deletes, let $v\in\mathit{Rel}(u)\setminus X_\Delta$,} and let $i$ be a repair test outside $F_\Delta\cup C_\Delta\cup B_\Delta$ whose trace covers $v$ but no instruction of $X_\Delta$.
\chong{Consider an execution agent whose outcome on each test $j$ depends only on the text of the instructions in $\mathit{Ctx}_t(j)$. It is one-hop local for $\Delta$, because $\Delta$ leaves the source text of every instruction outside $X_\Delta$ unchanged (Theorem~\ref{thm:iteration}(ii) and Edit locality). Yet $v\in\mathit{cov}(\tau_i)$ and, by symmetry, $u\in\mathit{Rel}(v)$, so $u\in\mathit{Ctx}_t(i)$, and the rewritten or deleted text of $u$ may change the outcome of $i$, although $i\notin L_0(S')$.}
\end{proof}

\begin{proof}[Proof of Theorem~\ref{thm:iteration}] If \textsc{Keep} applies, (i)--(iii) hold trivially.
Otherwise \textsc{Accept} gives $S_{t+1}\in\mathit{Top}$, so by \textsc{Screen} and \textsc{Cand}, $S_{t+1}$ is the materialization of some $\Delta$ with $I_t\vdash\Delta\ \mathsf{ok}$ and $\mathcal{C}(S_{t+1})$ is well formed, which is~(i).
For~(ii), the premises of Edit locality hold for $\Delta$.
By \chong{the typing of coordinated patches}, the mutations of $\Delta$ target distinct instructions; materialization rewrites or deletes the source span of each \chong{authored} target, or inserts at the end of an anchor's span; and the source map gives distinct instructions disjoint spans (Definition~\ref{def:sourcemap}(ii)).
Hence the nonempty edit intervals are pairwise disjoint, no insertion position lies strictly inside one of them, and $M_t(n)$ is an unedited fragment for every $n\notin X_\Delta$, so Edit locality applies.
Part~(iii) is the second premise of \textsc{Accept}, with the outcomes of $S_{t+1}$ carried over to the next iteration.
\end{proof}

\chong{The repair-boundary tests $B_\Delta$ are the tests that the repair model labels \juan{\texttt{boundary} to probe cases at the edge of the change}.}
The host deduplicates the selected repair tests and orders them according to the frozen repair split.
\chong{The repair model names source failures, control tests, and boundary tests, and the host validates these tests, determines the footprint $X_\Delta$, and computes $Q_\Delta$.}
\chong{Section~\ref{sec:overview-walkthrough} illustrates these tests for the running-example patch, and Appendix~\ref{app:d010-example} gives its provenance.
Patched skills with an empty $L(S')$ are removed before validation.
A patched skill is in group \textbf{A} if $R_{\mathrm{target}}>0$ and it has no regression; \textbf{B} if $R_{\mathrm{target}}>0$ and it has regressions but more fail-to-pass tests than regressions; \textbf{C} if $R_{\mathrm{target}}>0$ and it has no more fail-to-pass tests than regressions; and \textbf{D} if $R_{\mathrm{target}}=0$.
Patches in D are discarded, and the rest are ordered A before B before C.}
\chong{Within a group, patches are ordered by a larger excess of fail-to-pass tests over regressions, then by more fail-to-pass tests, then by fewer regressions.}
\chong{Remaining ranking ties are broken by the identifiers of the patched skills in the run log, in ascending lexicographic order.}
If several patched skills satisfy the acceptance rule, \chong{\method accepts the one with the highest validation pass rate.
Equal rates are resolved by their order in the shortlist $\mathit{Top}$.}
\chong{The initial skill is evaluated on all repair tests.}
An accepted patched skill becomes $S_{t+1}$.
If another iteration remains, \chong{that iteration starts by evaluating the new skill on all repair tests} to obtain its version-specific traces.
If the current skill is retained, its previously computed full-suite report is reused.
Earlier patch outcomes remain available to the repair model through node-linked repair memory.

\subsection{Case Study Provenance} \label{app:d010-example}

\chong{This subsection gives the identifiers behind the case study of Section~\ref{sec:case-study}.}
\juan{The turnover question is DocVQA question \texttt{15094}.
The patch is \texttt{r0-c1}, a single \texttt{MODIFY\_NODE} rewrite of instruction \texttt{S94e07a875c83510a}, labeled \texttt{D010} in the initial skill.
The supplied CDG context links the value/count/page condition \texttt{S079d0e6ec05dc1f4} to that instruction through \texttt{true\_control}, and the CFG links the edited instruction to \texttt{S414bf500cbf5fe15}, the similar-nearby-strings condition, through \texttt{sequential}.
The host uses the latter condition's saved coverage to include tests beyond those directly covering the edited instruction.}
\juan{The instruction, generation request and response, local comparison, and acceptance record \chong{belong to the DocVQA run recorded in the repair logs as \texttt{run-}\allowbreak\texttt{096f00aabdc74c16}\allowbreak\texttt{a4fa2150f24aa9fb}, which uses the GPT-6 Sol snapshot \texttt{gpt-6-sol-}\allowbreak\texttt{2026-09-22} as its execution model.}
Its saved evaluation records the complete turnover answer of Table~\ref{tab:turnover}.
The recorded skill hashes bind the repair and comparisons to the corresponding versions.}

\subsection{Patch Quality in Archived Trajectories} \label{app:archive}

\chong{Beyond the two repair runs of the repair mechanism analysis, we examine archived repair trajectories to compare patches before and after selection.}
\chong{This archive, which mixes trajectories from several versions of \method, contains 201 evaluated patches and 20 accepted patches from 13 trajectories with improved final test scores.}
Table~\ref{tab:candidate-quality} compares all evaluated patches with the subset of accepted patches.
The columns summarize sets of patches and their repair-test outcomes rather than paired before-and-after evaluations of the same tests.
Accepted patches more often fix \chong{the failing tests that motivated them.
\chong{Of the accepted patches, 95.0\% fix at least one source failure,} compared with 48.8\% of all evaluated patches, and they fix 68.4\% of their source failures,} compared with 40.6\% overall.
\chong{The archive includes one trajectory \chong{that accepted a group-D patch, whereas the current selection rule excludes} such patches before validation.}
Because \textsc{Screen} admits only patches that fix at least one target test, the higher target-repair rate of accepted patches partly follows from the selection rule, and because the archive keeps only trajectories whose final test scores improved, these rates describe successful runs rather than all runs.
At the same time, 94.6\% of previously passing repair tests still pass under accepted patches, \chong{which also keep 92.6\% and 93.3\% of their declared control and boundary tests passing.}
These controls are repair tests declared with the \texttt{preserve} or \texttt{boundary} role (Appendix~\ref{app:localization}).
Only evaluated controls that passed under the parent skill enter these rates.
\chong{The preservation rates differ less between the two sets than the repair rates (94.6\% versus 93.1\%, 92.6\% versus 87.5\%, and 93.3\% versus 92.8\%).}
\chong{}

\begin{table}[!htbp]
\centering
\caption{\chong{Repair-test outcomes (\%) for all 201 evaluated patches and the 20 accepted patches in 13 archived DocVQA and LiveMath repair trajectories with improved final test scores.}
Rates pool the corresponding counts across records.
Control rows include only declared and evaluated controls that
passed under the parent skill.}
\label{tab:candidate-quality}
\papertablestyle
\begin{tabularx}{\linewidth}{@{}>{\centering\arraybackslash}Xcc@{}}
\toprule
Measurement
& \shortstack{All patches}
& \shortstack{Accepted patches} \\
\midrule
Patches fixing at least one \chong{source failure}
& 48.8
& 95.0 \\

\chong{Source failures fixed}
& 40.6
& 68.4 \\

Previously passing repair tests still passing
& 93.1
& 94.6 \\

Declared \chong{control tests} still passing
& 87.5
& 92.6 \\

Declared boundary \chong{tests} still passing
& 92.8
& 93.3 \\

\bottomrule
\end{tabularx}
\end{table}

\end{document}